# Exact expressions of correlation functions between two spins in the boundary row of the two-dimensional rectangular Ising model with periodic-free boundary conditions and finite size

Tao Mei

Department of Journal, Central China Normal University, Wuhan, Hubei PRO, People's Republic of China

E-Mail: meitao@mail.ccnu.edu.cn

**Abstract:** We present exact expressions of correlation functions between two spins in the boundary row of the two-dimensional rectangular Ising model with periodic-free boundary conditions and finite size. Some properties of exact expressions obtained are discussed. The fundamental property of the Ising model that the long range order emerge as the temperature decrease is shown clearly; expressions of the correlation functions in the thermodynamic limit varies depending on the order of taking limits of two parameters $L$ and $N$; the impact of different sizes on correlation functions is illustrated with the aid of diagrams. Expressions of the correlation function discussed in this paper in the thermodynamic limit has been presented by previous researchers, and we prove that expressions obtained in this article is identical in form to expressions provided by previous researchers in the thermodynamic limit.

## 1 Introduction

There is a lot of literature on Ising model all along; the calculations of spin-spin correlation functions are important subject in the research of the two-dimensional Ising model, there is a lot of literature dealing with asymptotic expressions and properties of correlation functions in the thermodynamic limit, see, for example, Refs.[1, 2].

On the other hand, although asymptotic expressions of the partition function and correlation functions in the thermodynamic limit may be more convenient for practical calculations, it is self-evident that exact expressions of the partition function and correlation functions have different theoretical meanings.

In Refs.[4, 5], the method of spinor analysis given by Ref.[3] has been employed to calculate exact expressions of correlation functions $\left\langle \sigma_{l,1}\sigma_{l,1+Q} \right\rangle$ between two spins in a column of the two-dimensional rectangular Ising model with finite size. However, for the model with periodic-periodic boundary condition, which is illustrated in Fig.1, this method is effective only when $Q$ is a small number, that is to say, this method can only calculate $\left\langle \sigma_{1,1}\sigma_{1,2} \right\rangle$, $\left\langle \sigma_{1,1}\sigma_{1,3} \right\rangle$, $\left\langle \sigma_{1,1}\sigma_{1,4} \right\rangle$, $\cdots$, etc, effectively, which belong to short range order.

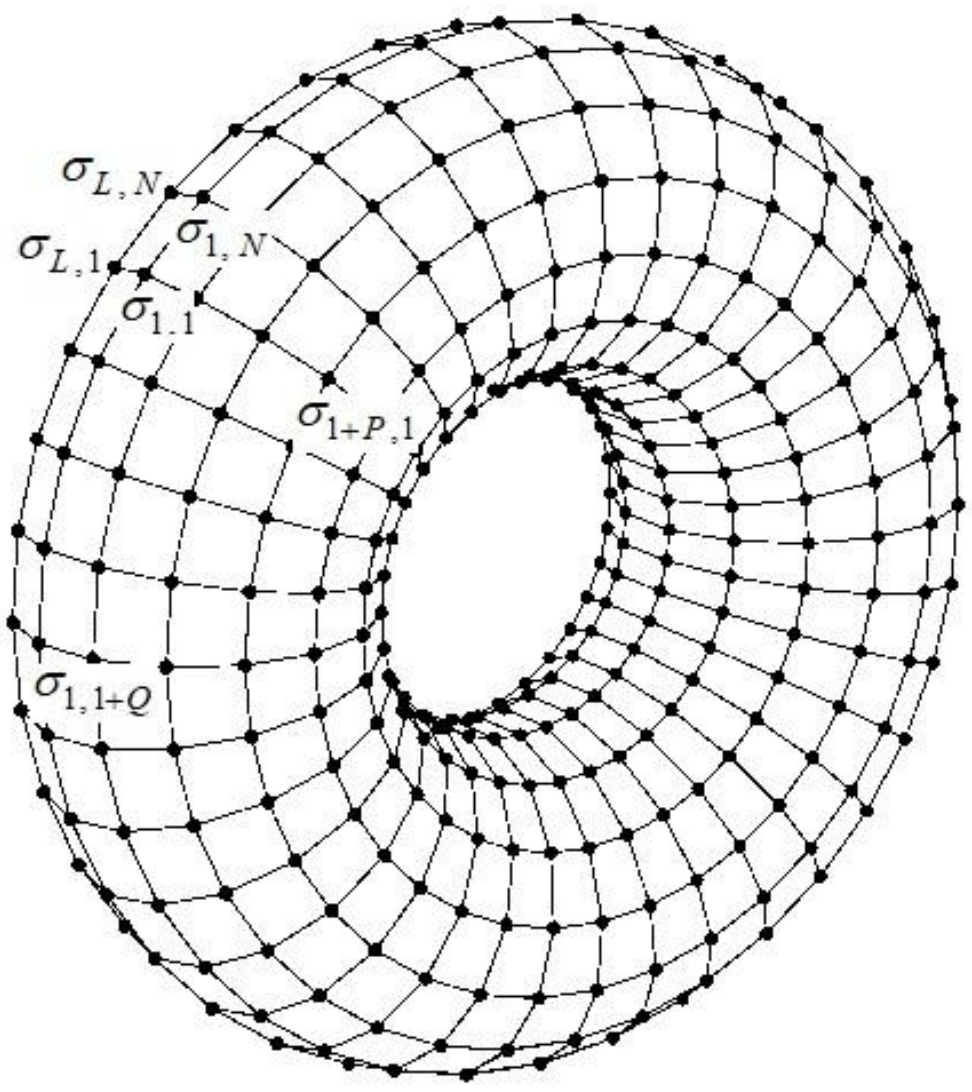

Fig.1 2-D Ising model on a torus

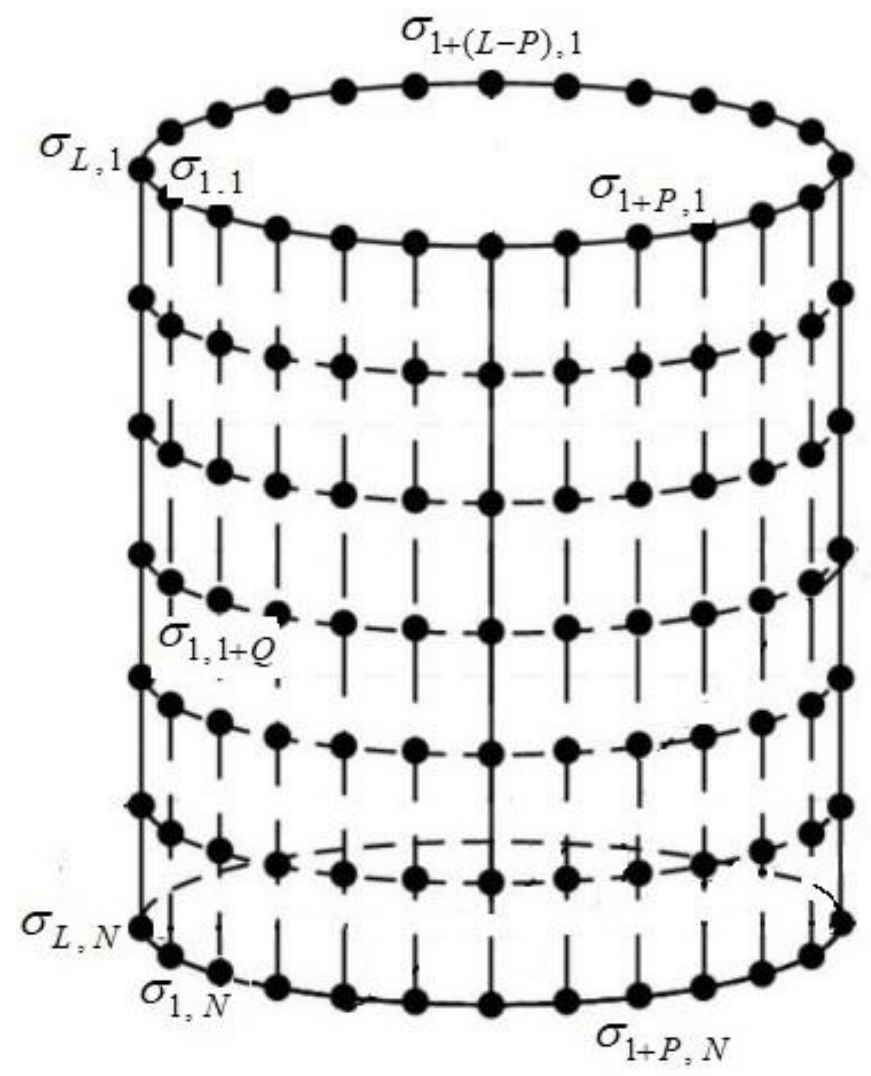

Fig.2 2-D Ising model on a cylinder

For the model with periodic-free boundary condition, which is illustrated in Fig.2, spinor analysis method is effective not only for calculating $\langle\sigma_{1,1}\sigma_{1,2}\rangle$, $\langle\sigma_{1,1}\sigma_{1,3}\rangle$, $\langle\sigma_{1,1}\sigma_{1,4}\rangle$, $\cdots$, but also for calculating $\langle\sigma_{1,1}\sigma_{1,N}\rangle$, $\langle\sigma_{1,1}\sigma_{1,N-1}\rangle$, $\langle\sigma_{1,1}\sigma_{1,N-2}\rangle,\cdots$, etc, which belong to long range order. From the exact expression of $\langle\sigma_{1,1}\sigma_{1,N}\rangle$ given in Ref.[5] it is shown clearly how the long range order emerge as the temperature decrease. However, this method still can not calculate $\langle\sigma_{1,1}\sigma_{1,1+Q}\rangle$ for all values of $Q=1,2,\cdots,N-1$.

In this paper, for two spins $\sigma_{1,1}$ and $\sigma_{1+P,1}$ in the boundary row of the two-dimensional rectangular Ising model with periodic-free boundary conditions, of which both positions are illustrated in Fig.2, we present exact expressions of correlation functions $\langle\sigma_{1,1}\sigma_{1+P,1}\rangle$ for all values of $P=1,2,\cdots,L-1$ in absence of magnetic field and finite size.

What we use in this paper is still the method of spinor analysis given by Ref.[3], since this method can only calculate exact expressions of the partition function of the model in absence of magnetic field, we only calculate $\langle\sigma_{1,1}\sigma_{1+P,1}\rangle$ in absence of magnetic field in this paper.

Since what we present are exact expressions of a whole row of correlation functions, which can reveal certain properties of correlation functions.

## 2 Definition, basic properties and the matrix form of $\langle\sigma_{1,1}\sigma_{1+P,1}\rangle$

The definitions of correlation functions between two spins in the boundary row of the two-dimensional rectangular Ising model with periodic-free boundary conditions in absence of magnetic field are

$$\left\langle \sigma_{1,1}\sigma_{1+P,1}\right\rangle = \frac{1}{Z}\sum_{\substack{\{\sigma_{h,v}=\pm1;\\ h=1,\cdots,L;\\ v=1,\cdots,N\}}} \sigma_{1,1}\sigma_{1+P,1}\exp\left(\frac{J'}{kT}\sum_{l'=1}^{L}\sum_{n'=1}^{N}\sigma_{l',n'}\sigma_{l'+1,n'}\right)\exp\left(\frac{J}{kT}\sum_{l=1}^{L}\sum_{n=1}^{N-1}\sigma_{l,n}\sigma_{l,n+1}\right), \quad (1)$$

and

$$Z = \sum_{\substack{\{\sigma_{h,v}=\pm1;\\ h=1,\cdots,L;\\ v=1,\cdots,N\}}} \exp\left(\frac{J'}{kT}\sum_{l=1}^{L}\sum_{n=1}^{N}\sigma_{l,n}\sigma_{l+1,n}\right)\exp\left(\frac{J}{kT}\sum_{l=1}^{L}\sum_{n=1}^{N-1}\sigma_{l,n}\sigma_{l,n+1}\right). \quad (2)$$

is the partition function of the model with periodic-free boundary condition in absence of magnetic field. In (1) and (2), $\sigma_{L+1,n}=\sigma_{1,n}$, $J'(>0)$ and $J(>0)$ are the interaction constants for the horizontal and vertical directions, respectively. As can be seen from (2), the model is with periodic boundary condition for the horizontal directions and with free boundary condition for the vertical directions, respectively.

Since the model is with periodic boundary condition for the horizontal directions, we have

$$\left\langle \sigma_{l,1}\sigma_{l+P,1}\right\rangle = \left\langle \sigma_{1,1}\sigma_{1+P,1}\right\rangle \qquad (l=1,2,\cdots,L).$$

And, since $N$-th row is also boundary row, by re-denoting $\sigma_{l,n}$ as $\sigma_{l,N-(n-1)}$ and considering periodic boundary condition for the horizontal directions, we have

$$\left\langle \sigma_{l,N}\sigma_{l+P,N}\right\rangle = \left\langle \sigma_{1,N}\sigma_{1+P,N}\right\rangle = \left\langle \sigma_{1,1}\sigma_{1+P,1}\right\rangle \qquad (l=1,2,\cdots,L).$$

Because of periodic boundary condition for the horizontal directions, both $\sigma_{2,1}$ and $\sigma_{L,1}$ are the closest spins of $\sigma_{1,1}$, we have $\left\langle \sigma_{1,1}\sigma_{2,1}\right\rangle = \left\langle \sigma_{1,1}\sigma_{L,1}\right\rangle$; similarly, both $\sigma_{3,1}$ and $\sigma_{L-1,1}$ are the next closest spins of $\sigma_{1,1}$, we have $\left\langle \sigma_{1,1}\sigma_{3,1}\right\rangle = \left\langle \sigma_{1,1}\sigma_{L-1,1}\right\rangle$; $\cdots\cdots$. Generally speaking, according to periodic boundary condition for the horizontal directions, it is easy to prove

$$\left\langle \sigma_{1,1}\sigma_{1+P,1}\right\rangle = \left\langle \sigma_{1,1}\sigma_{1+(L-P),1}\right\rangle. \quad (3)$$

As can be seen from Fig.2, the above three properties are obvious.

Hence, $\sigma_{1+[L/2],1}$ is the farthest spin of $\sigma_{1,1}$, where $[x]$ denotes the greatest integer not exceeding $x$; therefore, $\left\langle \sigma_{1,1}\sigma_{1+[L/2],1}\right\rangle$ is the "longest range order" in the boundary row of the model, and, further, when $P=Q_1$ or $P=L-Q_1$, $Q_1=1,2,3,\cdots$ be small positive integer, $\left\langle \sigma_{1,1}\sigma_{1+P,1}\right\rangle$ belong to short range order; when $P=\left[\frac{L}{2}\right]\pm Q_2$, $Q_2=0,1,2,3,\cdots$ be zero or small positive integer, $\left\langle \sigma_{1,1}\sigma_{1+P,1}\right\rangle$ belong to long range order.

$\left\langle \sigma_{1,1}\sigma_{1+P,1}\right\rangle$ given by (1) can be expressed as the matrix form:

$$\left\langle \sigma_{1,1}\sigma_{1+P,1}\right\rangle = \frac{1}{Z}\left(2\sinh\frac{2J'}{kT}\right)^{\frac{LN}{2}} \mathrm{Tr}\left(\mathbf{Z}_1(\mathbf{V}_2\mathbf{V}_1)^P \mathbf{Z}_1(\mathbf{V}_2\mathbf{V}_1)^{L-P}\right), \tag{4}$$

where matrices $\mathbf{V}_1$, $\mathbf{V}_2$ and $\mathbf{Z}_1$ are given by (A3) and (A6), respectively.

In this paper, for example, (A1) and (B2) represent formulas numbered (A1) and (B2) in Appendix A and Appendix B, respectively.

As can be seen from concrete forms of the three matrixes $\mathbf{Z}_1$, $\mathbf{\Gamma}_1$ and $\mathbf{V}_2^{1/2}$ given by (A6), (A5) and (A3), respectively, $\mathbf{Z}_1 = \mathbf{\Gamma}_1$, and both $\mathbf{\Gamma}_1$ and $\mathbf{V}_2^{1/2}$ are diagonal matrices, hence, $\left[\mathbf{\Gamma}_1, \mathbf{V}_2^{1/2}\right] = 0$, in terms of these two results, by simple conclusions, we can write the term $\mathrm{Tr}\left(\mathbf{Z}_1(\mathbf{V}_2\mathbf{V}_1)^P \mathbf{Z}_1(\mathbf{V}_2\mathbf{V}_1)^{L-P}\right)$ in (4) in the form

$$\mathrm{Tr}\left(\mathbf{Z}_1(\mathbf{V}_2\mathbf{V}_1)^P \mathbf{Z}_1(\mathbf{V}_2\mathbf{V}_1)^{L-P}\right) = \mathrm{Tr}\left(\mathbf{\Gamma}_1\left(\mathbf{V}_2^{1/2}\mathbf{V}_1\mathbf{V}_2^{1/2}\right)^P \mathbf{\Gamma}_1\left(\mathbf{V}_2^{1/2}\mathbf{V}_1\mathbf{V}_2^{1/2}\right)^{L-P}\right). \tag{5}$$

In terms of $[\mathbf{A}, \mathbf{B}] = \mathbf{AB} - \mathbf{BA}$, the term $\mathbf{\Gamma}_1\left(\mathbf{V}_2^{1/2}\mathbf{V}_1\mathbf{V}_2^{1/2}\right)^P \mathbf{\Gamma}_1$ in (5) can be written in the form

$$\mathbf{\Gamma}_1\left(\mathbf{V}_2^{1/2}\mathbf{V}_1\mathbf{V}_2^{1/2}\right)^P \mathbf{\Gamma}_1 = \left(\mathbf{V}_2^{1/2}\mathbf{V}_1\mathbf{V}_2^{1/2}\right)^P - \frac{1}{2}\left[\mathbf{\Gamma}_1, \left[\mathbf{\Gamma}_1, \left(\mathbf{V}_2^{1/2}\mathbf{V}_1\mathbf{V}_2^{1/2}\right)^P\right]\right],$$

substituting the above expression into (5) and according to the properties of trace: $\mathrm{Tr}(\mathbf{A}+\mathbf{B}) = \mathrm{Tr}\mathbf{A} + \mathrm{Tr}\mathbf{B}$ and $\mathrm{Tr}(-\mathbf{A}) = -\mathrm{Tr}\mathbf{A}$, we obtain

$$\begin{aligned}
&\mathrm{Tr}\left(\mathbf{Z}_1(\mathbf{V}_2\mathbf{V}_1)^P \mathbf{Z}_1(\mathbf{V}_2\mathbf{V}_1)^{L-P}\right) \\
&= \mathrm{Tr}\left(\left(\left(\mathbf{V}_2^{1/2}\mathbf{V}_1\mathbf{V}_2^{1/2}\right)^P - \frac{1}{2}\left[\mathbf{\Gamma}_1, \left[\mathbf{\Gamma}_1, \left(\mathbf{V}_2^{1/2}\mathbf{V}_1\mathbf{V}_2^{1/2}\right)^P\right]\right]\right)\left(\mathbf{V}_2^{1/2}\mathbf{V}_1\mathbf{V}_2^{1/2}\right)^{L-P}\right) \\
&= \mathrm{Tr}\left(\left(\mathbf{V}_2^{1/2}\mathbf{V}_1\mathbf{V}_2^{1/2}\right)^L - \frac{1}{2}\left[\mathbf{\Gamma}_1, \left[\mathbf{\Gamma}_1, \left(\mathbf{V}_2^{1/2}\mathbf{V}_1\mathbf{V}_2^{1/2}\right)^P\right]\right]\left(\mathbf{V}_2^{1/2}\mathbf{V}_1\mathbf{V}_2^{1/2}\right)^{L-P}\right) \\
&= \mathrm{Tr}\left(\left(\mathbf{V}_2^{1/2}\mathbf{V}_1\mathbf{V}_2^{1/2}\right)^L\right) - \frac{1}{2}\mathrm{Tr}\left(\left[\mathbf{\Gamma}_1, \left[\mathbf{\Gamma}_1, \left(\mathbf{V}_2^{1/2}\mathbf{V}_1\mathbf{V}_2^{1/2}\right)^P\right]\right]\left(\mathbf{V}_2^{1/2}\mathbf{V}_1\mathbf{V}_2^{1/2}\right)^{L-P}\right) \\
&= Z_0 - \frac{1}{2}\mathrm{Tr}\left(\left[\mathbf{\Gamma}_1, \left[\mathbf{\Gamma}_1, \left(\mathbf{V}_2^{1/2}\mathbf{V}_1\mathbf{V}_2^{1/2}\right)^P\right]\right]\left(\mathbf{V}_2^{1/2}\mathbf{V}_1\mathbf{V}_2^{1/2}\right)^{L-P}\right),
\end{aligned} \tag{6}$$

in the last step of the calculation process in the above formula, we have used the expression of $Z_0$ given by (A2).

Substituting the partition function $Z$ given by (A1) and the result given in (6) into (4), we obtain

$$\left\langle \sigma_{1,1}\sigma_{1+P,1}\right\rangle = 1 - \frac{1}{2Z_0}\mathrm{Tr}\left(\left[\mathbf{\Gamma}_1, \left[\mathbf{\Gamma}_1, \left(\mathbf{V}_2^{1/2}\mathbf{V}_1\mathbf{V}_2^{1/2}\right)^P\right]\right]\left(\mathbf{V}_2^{1/2}\mathbf{V}_1\mathbf{V}_2^{1/2}\right)^{L-P}\right). \tag{7}$$

where the matrix $\mathbf{\Gamma}_1$ is given by (A5).

The main purpose of this paper is to calculate (7), what method we employ to calculate exact expressions of (7) is only straightforward calculation based on some results given in Refs.[4, 5], which are listed in Appendix A of this paper.

## 3 The calculation of the second term in Eq.(7)

The calculation of the second term in (7) is complex and involves many steps.

(1) Since (A34) holds for arbitrary positive integer $L$, according to (A34) we have

$$\left(\mathbf{V}_2^{1/2}\mathbf{V}_1\mathbf{V}_2^{1/2}\right)^{L-P} = R(\xi)\left(\prod_{n=1}^{N}\mathrm{e}^{\frac{(L-P)\gamma_{n-1}}{2}\mathbf{\Gamma}_{2n-1}\mathrm{i}\mathbf{\Gamma}_{2n}}\right)R\left(\xi^{\mathrm{T}}\right), \tag{8}$$

according to (A34) and (A30) we have

$$\left(\mathbf{V}_2^{1/2}\mathbf{V}_1\mathbf{V}_2^{1/2}\right)^{P} = R(\xi)\left(\prod_{n=1}^{N}\mathrm{e}^{\frac{P\gamma_{n-1}}{2}\mathbf{\Gamma}_{2n-1}\mathrm{i}\mathbf{\Gamma}_{2n}}\right)R\left(\xi^{\mathrm{T}}\right) = \prod_{n=1}^{N}\left(R(\xi)\mathrm{e}^{\frac{P\gamma_{n-1}}{2}\mathbf{\Gamma}_{2n-1}\mathrm{i}\mathbf{\Gamma}_{2n}}R\left(\xi^{\mathrm{T}}\right)\right). \tag{9}$$

In terms of[3]

$$\mathrm{e}^{\frac{\theta}{2}\mathbf{\Gamma}_\mu\mathbf{\Gamma}_\nu} = \cos\frac{\theta}{2} + \mathbf{\Gamma}_\mu\mathbf{\Gamma}_\nu\sin\frac{\theta}{2}, \tag{10}$$

we have

$$\begin{aligned} R(\xi)\mathrm{e}^{\frac{P\gamma_{n-1}}{2}\mathbf{\Gamma}_{2n-1}\mathrm{i}\mathbf{\Gamma}_{2n}}R\left(\xi^{\mathrm{T}}\right) &= R(\xi)\cosh\frac{P\gamma_{n-1}}{2}R\left(\xi^{\mathrm{T}}\right) + R(\xi)\mathbf{\Gamma}_{2n-1}R\left(\xi^{\mathrm{T}}\right)\mathrm{i}R(\xi)\mathbf{\Gamma}_{2n}R\left(\xi^{\mathrm{T}}\right)\sinh\frac{P\gamma_{n-1}}{2} \\ &= \cosh\frac{P\gamma_{n-1}}{2} + \widetilde{\mathbf{\Gamma}}_{2n-1}\mathrm{i}\widetilde{\mathbf{\Gamma}}_{2n}\sinh\frac{P\gamma_{n-1}}{2} = \mathrm{e}^{\frac{P\gamma_{n-1}}{2}\widetilde{\mathbf{\Gamma}}_{2n-1}\mathrm{i}\widetilde{\mathbf{\Gamma}}_{2n}}. \end{aligned} \tag{11}$$

Substituting the result given by (11) into (9), we obtain

$$\left(\mathbf{V}_2^{1/2}\mathbf{V}_1\mathbf{V}_2^{1/2}\right)^{P} = \prod_{n=1}^{N}\mathrm{e}^{\frac{P\gamma_{n-1}}{2}\widetilde{\mathbf{\Gamma}}_{2n-1}\mathrm{i}\widetilde{\mathbf{\Gamma}}_{2n}}. \tag{12}$$

Spin representative has the property[3]: $R\left(\boldsymbol{\mu}^{\mathrm{T}}\right)\mathbf{\Gamma}_n R(\boldsymbol{\mu}) = \sum_{n'}^{2N}\mu_{nn'}\mathbf{\Gamma}_{n'}$, where $\mu_{nn'}$ are the elements of orthogonal matrix $\boldsymbol{\mu}$, when $\boldsymbol{\mu} = \xi$, in terms of (A23), the matrixes $\widetilde{\mathbf{\Gamma}}_n$ $(n = 1, 2, \cdots, 2N)$ introduced in (11) read

$$\widetilde{\mathbf{\Gamma}}_{2n-1} = R(\xi)\mathbf{\Gamma}_{2n-1}R\left(\xi^{\mathrm{T}}\right) = \sum_{p=1}^{2N}\left(\xi^{\mathrm{T}}\right)_{2n-1,\,p}\mathbf{\Gamma}_p = \sum_{p=1}^{2N}(\xi)_{p,\,2n-1}\mathbf{\Gamma}_p = \sum_{p=1}^{N}u_{pn}\mathbf{\Gamma}_{2p-1}, \tag{13}$$

$$\widetilde{\mathbf{\Gamma}}_{2n} = R(\xi)\mathbf{\Gamma}_{2n}R\left(\xi^{\mathrm{T}}\right) = \sum_{q=1}^{2N}\left(\xi^{\mathrm{T}}\right)_{2n,\,q}\mathbf{\Gamma}_q = \sum_{q=1}^{2N}(\xi)_{q,\,2n}\mathbf{\Gamma}_q = \sum_{q=1}^{N}v_{qn}\mathbf{\Gamma}_{2q}. \tag{14}$$

According to (A8) and (A30), it is easy to prove that the matrixes $\widetilde{\mathbf{\Gamma}}_n$ $(n = 1, 2, \cdots, 2N)$ given as (13) and (14) satisfy

$$\left\{\widetilde{\mathbf{\Gamma}}_l, \widetilde{\mathbf{\Gamma}}_m\right\} = 2\delta_{lm} \qquad (l, m = 1, 2, \cdots, 2N). \tag{15}$$

(2) For $\left(\mathbf{V}_2^{1/2}\mathbf{V}_1\mathbf{V}_2^{1/2}\right)^{P}$ given by (12), using $[A, CD] = [A, C]D + C[A, D]$ we have

$$\left[\boldsymbol{\Gamma}_1,\left(\mathbf{V}_2^{1/2}\mathbf{V}_1\mathbf{V}_2^{1/2}\right)^P\right]=\left[\boldsymbol{\Gamma}_1,\prod_{n=1}^{N}\mathrm{e}^{\frac{P\gamma_{n-1}}{2}\tilde{\boldsymbol{\Gamma}}_{2n-1}\mathrm{i}\tilde{\boldsymbol{\Gamma}}_{2n}}\right]$$

$$=\sum_{k=1}^{N}\left(\prod_{n=1}^{k-1}\mathrm{e}^{\frac{P\gamma_{n-1}}{2}\tilde{\boldsymbol{\Gamma}}_{2n-1}\mathrm{i}\tilde{\boldsymbol{\Gamma}}_{2n}}\right)\left[\boldsymbol{\Gamma}_1,\mathrm{e}^{\frac{P\gamma_{k-1}}{2}\tilde{\boldsymbol{\Gamma}}_{2k-1}\mathrm{i}\tilde{\boldsymbol{\Gamma}}_{2k}}\right]\left(\prod_{n=k+1}^{N}\mathrm{e}^{\frac{P\gamma_{n-1}}{2}\tilde{\boldsymbol{\Gamma}}_{2n-1}\mathrm{i}\tilde{\boldsymbol{\Gamma}}_{2n}}\right). \tag{16}$$

For the term $\left[\boldsymbol{\Gamma}_1,\mathrm{e}^{\frac{P\gamma_{k-1}}{2}\tilde{\boldsymbol{\Gamma}}_{2k-1}\mathrm{i}\tilde{\boldsymbol{\Gamma}}_{2k}}\right]$ in (16), according to (11), (13) and (14) we have

$$\left[\boldsymbol{\Gamma}_1,\mathrm{e}^{\frac{P\gamma_{k-1}}{2}\tilde{\boldsymbol{\Gamma}}_{2k-1}\mathrm{i}\tilde{\boldsymbol{\Gamma}}_{2k}}\right]=\left[\boldsymbol{\Gamma}_1,\cosh\frac{P\gamma_{k-1}}{2}+\tilde{\boldsymbol{\Gamma}}_{2k-1}\mathrm{i}\tilde{\boldsymbol{\Gamma}}_{2k}\sinh\frac{P\gamma_{k-1}}{2}\right]=\mathrm{i}\sinh\frac{P\gamma_{k-1}}{2}\left[\boldsymbol{\Gamma}_1,\tilde{\boldsymbol{\Gamma}}_{2k-1}\mathrm{i}\tilde{\boldsymbol{\Gamma}}_{2k}\right]$$

$$=\mathrm{i}\sinh\frac{P\gamma_{k-1}}{2}\left[\boldsymbol{\Gamma}_1,\left(\sum_{p=1}^{N}u_{pk}\boldsymbol{\Gamma}_{2p-1}\right)\left(\sum_{q=1}^{N}v_{qk}\boldsymbol{\Gamma}_{2q}\right)\right]=\mathrm{i}\sinh\frac{P\gamma_{k-1}}{2}\sum_{p=1}^{N}u_{pk}\sum_{q=1}^{N}v_{qk}\left[\boldsymbol{\Gamma}_1,\boldsymbol{\Gamma}_{2p-1}\boldsymbol{\Gamma}_{2q}\right], \tag{17}$$

for the term $\left[\boldsymbol{\Gamma}_1,\boldsymbol{\Gamma}_{2p-1}\boldsymbol{\Gamma}_{2q}\right]$ in (17), according to (A8) and $[A,CD]=\{A,C\}D-C\{A,D\}$ we obtain

$$\left[\boldsymbol{\Gamma}_1,\boldsymbol{\Gamma}_{2p-1}\boldsymbol{\Gamma}_{2q}\right]=\left\{\boldsymbol{\Gamma}_1,\boldsymbol{\Gamma}_{2p-1}\right\}\boldsymbol{\Gamma}_{2q}-\boldsymbol{\Gamma}_{2p-1}\left\{\boldsymbol{\Gamma}_1,\boldsymbol{\Gamma}_{2q}\right\}=2\delta_{1p}\boldsymbol{\Gamma}_{2q},$$

substituting the above result into (17) and in terms of (14) we obtain

$$\left[\boldsymbol{\Gamma}_1,\mathrm{e}^{\frac{P\gamma_{k-1}}{2}\tilde{\boldsymbol{\Gamma}}_{2k-1}\mathrm{i}\tilde{\boldsymbol{\Gamma}}_{2k}}\right]=2\mathrm{i}\sinh\frac{P\gamma_{k-1}}{2}u_{1k}\sum_{q=1}^{N}v_{qk}\boldsymbol{\Gamma}_{2q}=2\sinh\frac{P\gamma_{k-1}}{2}u_{1k}\mathrm{i}\tilde{\boldsymbol{\Gamma}}_{2k}.$$

Substituting the above result into (16), we have

$$\left[\boldsymbol{\Gamma}_1,\left(\mathbf{V}_2^{1/2}\mathbf{V}_1\mathbf{V}_2^{1/2}\right)^P\right]=2\sum_{k=1}^{N}\sinh\frac{P\gamma_{k-1}}{2}u_{1k}\left(\prod_{n=1}^{k-1}\mathrm{e}^{\frac{P\gamma_{n-1}}{2}\tilde{\boldsymbol{\Gamma}}_{2n-1}\mathrm{i}\tilde{\boldsymbol{\Gamma}}_{2n}}\right)\mathrm{i}\tilde{\boldsymbol{\Gamma}}_{2k}\left(\prod_{n=k+1}^{N}\mathrm{e}^{\frac{P\gamma_{n-1}}{2}\tilde{\boldsymbol{\Gamma}}_{2n-1}\mathrm{i}\tilde{\boldsymbol{\Gamma}}_{2n}}\right),$$

according to $\left[\tilde{\boldsymbol{\Gamma}}_{2k},\tilde{\boldsymbol{\Gamma}}_{2n-1}\tilde{\boldsymbol{\Gamma}}_{2n}\right]=0 \quad (n\neq k)$ obtained by (15) we obtain

$$\left[\boldsymbol{\Gamma}_1,\left(\mathbf{V}_2^{1/2}\mathbf{V}_1\mathbf{V}_2^{1/2}\right)^P\right]=2\sum_{m=1}^{N}u_{1m}\sinh\frac{P\gamma_{m-1}}{2}\mathrm{i}\tilde{\boldsymbol{\Gamma}}_{2m}\left(\prod_{\substack{n=1\\n\neq m}}^{N}\mathrm{e}^{\frac{P\gamma_{n-1}}{2}\tilde{\boldsymbol{\Gamma}}_{2n-1}\mathrm{i}\tilde{\boldsymbol{\Gamma}}_{2n}}\right). \tag{18}$$

(3) Using the result given in (18) we have

$$\left[\boldsymbol{\Gamma}_1,\left[\boldsymbol{\Gamma}_1,\left(\mathbf{V}_2^{1/2}\mathbf{V}_1\mathbf{V}_2^{1/2}\right)^P\right]\right]=2\sum_{m=1}^{N}u_{1m}\sinh\frac{P\gamma_{m-1}}{2}\mathrm{i}\left[\boldsymbol{\Gamma}_1,\tilde{\boldsymbol{\Gamma}}_{2m}\left(\prod_{\substack{n=1\\n\neq m}}^{N}\mathrm{e}^{\frac{P\gamma_{n-1}}{2}\tilde{\boldsymbol{\Gamma}}_{2n-1}\mathrm{i}\tilde{\boldsymbol{\Gamma}}_{2n}}\right)\right]=\mathbf{W}_1+\mathbf{W}_2, \tag{19}$$

$$\mathbf{W}_1=2\sum_{m=1}^{N}u_{1m}\sinh\frac{P\gamma_{m-1}}{2}\mathrm{i}\left[\boldsymbol{\Gamma}_1,\tilde{\boldsymbol{\Gamma}}_{2m}\right]\left(\prod_{\substack{n=1\\n\neq m}}^{N}\mathrm{e}^{\frac{P\gamma_{n-1}}{2}\tilde{\boldsymbol{\Gamma}}_{2n-1}\mathrm{i}\tilde{\boldsymbol{\Gamma}}_{2n}}\right), \tag{20}$$

$$\mathbf{W}_2=2\sum_{m=1}^{N}u_{1m}\sinh\frac{P\gamma_{m-1}}{2}\mathrm{i}\tilde{\boldsymbol{\Gamma}}_{2m}\left[\boldsymbol{\Gamma}_1,\prod_{\substack{n=1\\n\neq m}}^{N}\mathrm{e}^{\frac{P\gamma_{n-1}}{2}\tilde{\boldsymbol{\Gamma}}_{2n-1}\mathrm{i}\tilde{\boldsymbol{\Gamma}}_{2n}}\right]. \tag{21}$$

For the term $\left[\boldsymbol{\Gamma}_1, \tilde{\boldsymbol{\Gamma}}_{2m}\right]$ in $\mathbf{W}_1$ given by (20), according to (A8) and (14), we have

$$\left[\boldsymbol{\Gamma}_1, \tilde{\boldsymbol{\Gamma}}_{2m}\right] = \left[\boldsymbol{\Gamma}_1, \sum_{q=1}^{N} v_{qm}\boldsymbol{\Gamma}_{2q}\right] = \sum_{q=1}^{N} v_{qm}\left[\boldsymbol{\Gamma}_1, \boldsymbol{\Gamma}_{2q}\right] = \sum_{q=1}^{N} 2v_{qm}\boldsymbol{\Gamma}_1\boldsymbol{\Gamma}_{2q} = 2\boldsymbol{\Gamma}_1\tilde{\boldsymbol{\Gamma}}_{2m},$$

substituting the above result into (20), in terms of (A30), (11), (13) and (14), we obtain

$$\begin{aligned}
\mathbf{W}_1 &= 4\sum_{m=1}^{N} u_{1m}\sinh\frac{P\gamma_{m-1}}{2}\,\mathrm{i}\boldsymbol{\Gamma}_1\tilde{\boldsymbol{\Gamma}}_{2m}\left(\prod_{\substack{n=1\\ n\neq m}}^{N} \mathrm{e}^{\frac{P\gamma_{n-1}}{2}\tilde{\boldsymbol{\Gamma}}_{2n-1}\mathrm{i}\tilde{\boldsymbol{\Gamma}}_{2n}}\right)\\
&= 4\sum_{m=1}^{N} u_{1m}\sinh\frac{P\gamma_{m-1}}{2}\,\mathrm{i}\boldsymbol{\Gamma}_1 R(\boldsymbol{\xi})\boldsymbol{\Gamma}_{2m}R\left(\boldsymbol{\xi}^{\mathrm{T}}\right)\left(\prod_{\substack{n=1\\ n\neq m}}^{N} R(\boldsymbol{\xi})\mathrm{e}^{\frac{P\gamma_{n-1}}{2}\boldsymbol{\Gamma}_{2n-1}\mathrm{i}\boldsymbol{\Gamma}_{2n}}R\left(\boldsymbol{\xi}^{\mathrm{T}}\right)\right)\\
&= 4\sum_{m=1}^{N} u_{1m}\sinh\frac{P\gamma_{m-1}}{2}\,\boldsymbol{\Gamma}_1 R(\boldsymbol{\xi})\mathrm{i}\boldsymbol{\Gamma}_{2m}\left(\prod_{\substack{n=1\\ n\neq m}}^{N} \mathrm{e}^{\frac{P\gamma_{n-1}}{2}\boldsymbol{\Gamma}_{2n-1}\mathrm{i}\boldsymbol{\Gamma}_{2n}}\right)R\left(\boldsymbol{\xi}^{\mathrm{T}}\right).
\end{aligned} \tag{22}$$

As for $\mathbf{W}_2$ given by (21), we first employ the derivation process similar to that from (16) to (18) and obtain

$$\left[\boldsymbol{\Gamma}_1, \prod_{\substack{n=1\\ n\neq m}}^{N} \mathrm{e}^{\frac{P\gamma_{n-1}}{2}\tilde{\boldsymbol{\Gamma}}_{2n-1}\mathrm{i}\tilde{\boldsymbol{\Gamma}}_{2n}}\right] = 2\sum_{\substack{l=1\\ l\neq m}}^{N} u_{1l}\sinh\frac{P\gamma_{l-1}}{2}\,\mathrm{i}\tilde{\boldsymbol{\Gamma}}_{2l}\left(\prod_{\substack{n=1\\ n\neq m,\, n\neq l}}^{N} \mathrm{e}^{\frac{P\gamma_{n-1}}{2}\tilde{\boldsymbol{\Gamma}}_{2n-1}\mathrm{i}\tilde{\boldsymbol{\Gamma}}_{2n}}\right),$$

substituting the above result into (21), and using (15), we obtain

$$\begin{aligned}
\mathbf{W}_2 &= 2\sum_{m=1}^{N} u_{1m}\sinh\frac{P\gamma_{m-1}}{2}\,\mathrm{i}\tilde{\boldsymbol{\Gamma}}_{2m}\left(2\sum_{\substack{l=1\\ l\neq m}}^{N} u_{1l}\sinh\frac{P\gamma_{l-1}}{2}\,\mathrm{i}\tilde{\boldsymbol{\Gamma}}_{2l}\left(\prod_{\substack{n=1\\ n\neq m,\, n\neq l}}^{N} \mathrm{e}^{\frac{P\gamma_{n-1}}{2}\tilde{\boldsymbol{\Gamma}}_{2n-1}\mathrm{i}\tilde{\boldsymbol{\Gamma}}_{2n}}\right)\right)\\
&= -4\sum_{m=1}^{N}\sum_{\substack{l=1\\ l>m}}^{N} u_{1m}u_{1l}\sinh\frac{P\gamma_{m-1}}{2}\sinh\frac{P\gamma_{l-1}}{2}\left\{\tilde{\boldsymbol{\Gamma}}_{2m}, \tilde{\boldsymbol{\Gamma}}_{2l}\right\}\left(\prod_{\substack{n=1\\ n\neq m,\, n\neq l}}^{N} \mathrm{e}^{\frac{P\gamma_{n-1}}{2}\tilde{\boldsymbol{\Gamma}}_{2n-1}\mathrm{i}\tilde{\boldsymbol{\Gamma}}_{2n}}\right)\\
&= 0.
\end{aligned} \tag{23}$$

Substituting the two results given in (22) and (23) into (19), we obtain

$$\left[\boldsymbol{\Gamma}_1, \left[\boldsymbol{\Gamma}_1, \left(\mathbf{V}_2^{1/2}\mathbf{V}_1\mathbf{V}_2^{1/2}\right)^P\right]\right] = 4\sum_{m=1}^{N} u_{1m}\sinh\frac{P\gamma_{m-1}}{2}\,\boldsymbol{\Gamma}_1 R(\boldsymbol{\xi})\mathrm{i}\boldsymbol{\Gamma}_{2m}\left(\prod_{\substack{n=1\\ n\neq m}}^{N} \mathrm{e}^{\frac{P\gamma_{n-1}}{2}\boldsymbol{\Gamma}_{2n-1}\mathrm{i}\boldsymbol{\Gamma}_{2n}}\right)R\left(\boldsymbol{\xi}^{\mathrm{T}}\right). \tag{24}$$

(4) According to two results in (8) and (24) and using (A30) we obtain

$$\left[\mathbf{\Gamma}_1,\left[\mathbf{\Gamma}_1,\left(\mathbf{V}_2^{1/2}\mathbf{V}_1\mathbf{V}_2^{1/2}\right)^P\right]\right]\left(\mathbf{V}_2^{1/2}\mathbf{V}_1\mathbf{V}_2^{1/2}\right)^{L-P}$$

$$=4\sum_{m=1}^{N}u_{1m}\sinh\frac{P\gamma_{m-1}}{2}\mathbf{\Gamma}_1 R(\xi)\mathrm{i}\mathbf{\Gamma}_{2m}\left(\prod_{\substack{n=1\\n\neq m}}^{N}\mathrm{e}^{\frac{P\gamma_{n-1}}{2}\mathbf{\Gamma}_{2n-1}\mathrm{i}\mathbf{\Gamma}_{2n}}\right)R\left(\xi^{\mathrm{T}}\right)\times R(\xi)\left(\prod_{n=1}^{N}\mathrm{e}^{\frac{(L-P)\gamma_{n-1}}{2}\mathbf{\Gamma}_{2n-1}\mathrm{i}\mathbf{\Gamma}_{2n}}\right)R\left(\xi^{\mathrm{T}}\right)$$

$$=4\sum_{m=1}^{N}u_{1m}\sinh\frac{P\gamma_{m-1}}{2}\mathbf{\Gamma}_1 R(\xi)\mathrm{i}\mathbf{\Gamma}_{2m}\left(\prod_{\substack{n=1\\n\neq m}}^{N}\mathrm{e}^{\frac{P\gamma_{n-1}}{2}\mathbf{\Gamma}_{2n-1}\mathrm{i}\mathbf{\Gamma}_{2n}}\right)\left(\prod_{n=1}^{N}\mathrm{e}^{\frac{(L-P)\gamma_{n-1}}{2}\mathbf{\Gamma}_{2n-1}\mathrm{i}\mathbf{\Gamma}_{2n}}\right)R\left(\xi^{\mathrm{T}}\right).$$

For above result, using $\mathrm{Tr}(\mathbf{A}+\mathbf{B})=\mathrm{Tr}(\mathbf{A})+\mathrm{Tr}\mathbf{B}$ and $\mathrm{Tr}(\mathbf{AB})=\mathrm{Tr}(\mathbf{BA})$ we obtain

$$\mathrm{Tr}\left(\left[\mathbf{\Gamma}_1,\left[\mathbf{\Gamma}_1,\left(\mathbf{V}_2^{1/2}\mathbf{V}_1\mathbf{V}_2^{1/2}\right)^P\right]\right]\left(\mathbf{V}_2^{1/2}\mathbf{V}_1\mathbf{V}_2^{1/2}\right)^{L-P}\right)$$

$$=\mathrm{Tr}\left(4\sum_{m=1}^{N}u_{1m}\sinh\frac{P\gamma_{m-1}}{2}\mathbf{\Gamma}_1 R(\xi)\mathrm{i}\mathbf{\Gamma}_{2m}\left(\prod_{\substack{n=1\\n\neq m}}^{N}\mathrm{e}^{\frac{P\gamma_{n-1}}{2}\mathbf{\Gamma}_{2n-1}\mathrm{i}\mathbf{\Gamma}_{2n}}\right)\left(\prod_{n=1}^{N}\mathrm{e}^{\frac{(L-P)\gamma_{n-1}}{2}\mathbf{\Gamma}_{2n-1}\mathrm{i}\mathbf{\Gamma}_{2n}}\right)R\left(\xi^{\mathrm{T}}\right)\right) \tag{25}$$

$$=4\sum_{m=1}^{N}u_{1m}\sinh\frac{P\gamma_{m-1}}{2}\mathrm{Tr}\left(R\left(\xi^{\mathrm{T}}\right)\mathbf{\Gamma}_1 R(\xi)\mathrm{i}\mathbf{\Gamma}_{2m}\left(\prod_{\substack{n=1\\n\neq m}}^{N}\mathrm{e}^{\frac{P\gamma_{n-1}}{2}\mathbf{\Gamma}_{2n-1}\mathrm{i}\mathbf{\Gamma}_{2n}}\right)\left(\prod_{n=1}^{N}\mathrm{e}^{\frac{(L-P)\gamma_{n-1}}{2}\mathbf{\Gamma}_{2n-1}\mathrm{i}\mathbf{\Gamma}_{2n}}\right)\right).$$

In terms of (A33),

$$\left(\prod_{\substack{n=1\\n\neq m}}^{N}\mathrm{e}^{\frac{P\gamma_{n-1}}{2}\mathbf{\Gamma}_{2n-1}\mathrm{i}\mathbf{\Gamma}_{2n}}\right)\left(\prod_{n=1}^{N}\mathrm{e}^{\frac{(L-P)\gamma_{n-1}}{2}\mathbf{\Gamma}_{2n-1}\mathrm{i}\mathbf{\Gamma}_{2n}}\right)=\mathrm{e}^{\frac{(L-P)\gamma_{m-1}}{2}\mathbf{\Gamma}_{2m-1}\mathrm{i}\mathbf{\Gamma}_{2m}}\left(\prod_{\substack{n=1\\n\neq m}}^{N}\mathrm{e}^{\frac{L\gamma_{n-1}}{2}\mathbf{\Gamma}_{2n-1}\mathrm{i}\mathbf{\Gamma}_{2n}}\right). \tag{26}$$

According to the property of spin representative: $R\left(\mathbf{\mu}^{\mathrm{T}}\right)\mathbf{\Gamma}_n R(\mathbf{\mu})=\sum_{n'}^{2N}\mu_{nn'}\mathbf{\Gamma}_{n'}$ , when $\mathbf{\mu}=\xi$ , in terms of (A23), we obtain

$$R\left(\xi^{\mathrm{T}}\right)\mathbf{\Gamma}_{2n-1}R(\xi)=\sum_{p=1}^{2N}(\xi)_{2n-1,\,p}\mathbf{\Gamma}_p=\sum_{p=1}^{N}u_{np}\mathbf{\Gamma}_{2p-1}\,, \tag{27}$$

notice that this result is different from (13).

From (27) we obtain $R\left(\xi^{\mathrm{T}}\right)\mathbf{\Gamma}_1 R(\xi)=\sum_{l=1}^{N}u_{1l}\mathbf{\Gamma}_{2l-1}$ , substituting this result and the result given by (26) into (25), and, in terms of $\mathrm{Tr}(\mathbf{A}+\mathbf{B})=\mathrm{Tr}(\mathbf{A})+\mathrm{Tr}\mathbf{B}$ , we obtain

$$\mathrm{Tr}\left(\left[\boldsymbol{\Gamma}_1,\left[\boldsymbol{\Gamma}_1,\left(\mathbf{V}_2^{1/2}\mathbf{V}_1\mathbf{V}_2^{1/2}\right)^P\right]\right]\left(\mathbf{V}_2^{1/2}\mathbf{V}_1\mathbf{V}_2^{1/2}\right)^{L-P}\right)$$

$$=4\sum_{m=1}^{N}u_{1m}\sinh\frac{P\gamma_{m-1}}{2}\sum_{l=1}^{N}u_{1l}\mathrm{Tr}\left(\boldsymbol{\Gamma}_{2l-1}\mathrm{i}\boldsymbol{\Gamma}_{2m}\mathrm{e}^{\frac{(L-P)\gamma_{m-1}}{2}\boldsymbol{\Gamma}_{2m-1}\mathrm{i}\boldsymbol{\Gamma}_{2m}}\left(\prod_{\substack{n=1\\n\neq m}}^{N}\mathrm{e}^{\frac{L\gamma_{n-1}}{2}\boldsymbol{\Gamma}_{2n-1}\mathrm{i}\boldsymbol{\Gamma}_{2n}}\right)\right)$$

$$=4\sum_{l=1}^{N}u_{1l}^2\sinh\frac{P\gamma_{l-1}}{2}\mathrm{Tr}\left(\boldsymbol{\Gamma}_{2l-1}\mathrm{i}\boldsymbol{\Gamma}_{2l}\mathrm{e}^{\frac{(L-P)\gamma_{l-1}}{2}\boldsymbol{\Gamma}_{2l-1}\mathrm{i}\boldsymbol{\Gamma}_{2l}}\left(\prod_{\substack{n=1\\n\neq l}}^{N}\mathrm{e}^{\frac{L\gamma_{n-1}}{2}\boldsymbol{\Gamma}_{2n-1}\mathrm{i}\boldsymbol{\Gamma}_{2n}}\right)\right) \tag{28}$$

$$+4\sum_{l=1}^{N}u_{1l}\sum_{\substack{m=1\\m\neq l}}^{N}u_{1m}\sinh\frac{P\gamma_{m-1}}{2}\mathrm{Tr}\left(\boldsymbol{\Gamma}_{2l-1}\mathrm{i}\boldsymbol{\Gamma}_{2m}\mathrm{e}^{\frac{(L-P)\gamma_{m-1}}{2}\boldsymbol{\Gamma}_{2m-1}\mathrm{i}\boldsymbol{\Gamma}_{2m}}\left(\prod_{\substack{n=1\\n\neq m}}^{N}\mathrm{e}^{\frac{L\gamma_{n-1}}{2}\boldsymbol{\Gamma}_{2n-1}\mathrm{i}\boldsymbol{\Gamma}_{2n}}\right)\right).$$

In terms of (A33), we have

$$\mathrm{e}^{\frac{(L-P)\gamma_{l-1}}{2}\boldsymbol{\Gamma}_{2l-1}\mathrm{i}\boldsymbol{\Gamma}_{2l}}\left(\prod_{\substack{n=1\\n\neq l}}^{N}\mathrm{e}^{\frac{L\gamma_{n-1}}{2}\boldsymbol{\Gamma}_{2n-1}\mathrm{i}\boldsymbol{\Gamma}_{2n}}\right)=\mathrm{e}^{\frac{(L-P)\gamma_{l-1}}{4}\boldsymbol{\Gamma}_{2l-1}\mathrm{i}\boldsymbol{\Gamma}_{2l}}\mathrm{e}^{\frac{(L-P)\gamma_{l-1}}{4}\boldsymbol{\Gamma}_{2l-1}\mathrm{i}\boldsymbol{\Gamma}_{2l}}\left(\prod_{\substack{n=1\\n\neq l}}^{N}\mathrm{e}^{\frac{L\gamma_{n-1}}{2}\boldsymbol{\Gamma}_{2n-1}\mathrm{i}\boldsymbol{\Gamma}_{2n}}\right)$$

$$=\mathrm{e}^{\frac{(L-P)\gamma_{l-1}}{4}\boldsymbol{\Gamma}_{2l-1}\mathrm{i}\boldsymbol{\Gamma}_{2l}}\left(\prod_{\substack{n=1\\n\neq l}}^{N}\mathrm{e}^{\frac{L\gamma_{n-1}}{2}\boldsymbol{\Gamma}_{2n-1}\mathrm{i}\boldsymbol{\Gamma}_{2n}}\right)\mathrm{e}^{\frac{(L-P)\gamma_{l-1}}{4}\boldsymbol{\Gamma}_{2l-1}\mathrm{i}\boldsymbol{\Gamma}_{2l}},$$

$$\mathrm{e}^{\frac{(L-P)\gamma_{m-1}}{2}\boldsymbol{\Gamma}_{2m-1}\mathrm{i}\boldsymbol{\Gamma}_{2m}}\left(\prod_{\substack{n=1\\n\neq m}}^{N}\mathrm{e}^{\frac{L\gamma_{n-1}}{2}\boldsymbol{\Gamma}_{2n-1}\mathrm{i}\boldsymbol{\Gamma}_{2n}}\right)$$

$$=\mathrm{e}^{\frac{(L-P)\gamma_{m-1}}{4}\boldsymbol{\Gamma}_{2m-1}\mathrm{i}\boldsymbol{\Gamma}_{2m}}\mathrm{e}^{\frac{(L-P)\gamma_{m-1}}{4}\boldsymbol{\Gamma}_{2m-1}\mathrm{i}\boldsymbol{\Gamma}_{2m}}\left(\prod_{\substack{n=1\\n\neq m}}^{l-1}\mathrm{e}^{\frac{L\gamma_{n-1}}{2}\boldsymbol{\Gamma}_{2n-1}\mathrm{i}\boldsymbol{\Gamma}_{2n}}\right)\mathrm{e}^{\frac{L\gamma_{l-1}}{4}\boldsymbol{\Gamma}_{2l-1}\mathrm{i}\boldsymbol{\Gamma}_{2l}}\mathrm{e}^{\frac{L\gamma_{l-1}}{4}\boldsymbol{\Gamma}_{2l-1}\mathrm{i}\boldsymbol{\Gamma}_{2l}}\left(\prod_{\substack{n=l+1\\n\neq m}}^{N}\mathrm{e}^{\frac{L\gamma_{n-1}}{2}\boldsymbol{\Gamma}_{2n-1}\mathrm{i}\boldsymbol{\Gamma}_{2n}}\right)$$

$$=\mathrm{e}^{\frac{L\gamma_{l-1}}{4}\boldsymbol{\Gamma}_{2l-1}\mathrm{i}\boldsymbol{\Gamma}_{2l}}\mathrm{e}^{\frac{(L-P)\gamma_{m-1}}{4}\boldsymbol{\Gamma}_{2m-1}\mathrm{i}\boldsymbol{\Gamma}_{2m}}\left(\prod_{\substack{n=1\\n\neq l,\,n\neq m}}^{N}\mathrm{e}^{\frac{L\gamma_{n-1}}{2}\boldsymbol{\Gamma}_{2n-1}\mathrm{i}\boldsymbol{\Gamma}_{2n}}\right)\mathrm{e}^{\frac{L\gamma_{l-1}}{4}\boldsymbol{\Gamma}_{2l-1}\mathrm{i}\boldsymbol{\Gamma}_{2l}}\mathrm{e}^{\frac{(L-P)\gamma_{m-1}}{4}\boldsymbol{\Gamma}_{2m-1}\mathrm{i}\boldsymbol{\Gamma}_{2m}}.$$

Substituting the above two results into (28), and, in terms of $\mathrm{Tr}(\mathbf{AB})=\mathrm{Tr}(\mathbf{BA})$, we obtain

$$\mathrm{Tr}\left(\left[\boldsymbol{\Gamma}_1,\left[\boldsymbol{\Gamma}_1,\left(\mathbf{V}_2^{1/2}\mathbf{V}_1\mathbf{V}_2^{1/2}\right)^P\right]\right]\left(\mathbf{V}_2^{1/2}\mathbf{V}_1\mathbf{V}_2^{1/2}\right)^{L-P}\right)$$

$$=4\sum_{l=1}^{N}u_{1l}^2\sinh\frac{P\gamma_{l-1}}{2}\mathrm{Tr}\mathbf{U}+4\sum_{l=1}^{N}u_{1l}\sum_{\substack{m=1\\m\neq l}}^{N}u_{1m}\sinh\frac{P\gamma_{m-1}}{2}\mathrm{Tr}\mathbf{V}, \tag{29}$$

$$\mathbf{U}=\mathrm{e}^{\frac{(L-P)\gamma_{l-1}}{4}\boldsymbol{\Gamma}_{2l-1}\mathrm{i}\boldsymbol{\Gamma}_{2l}}\boldsymbol{\Gamma}_{2l-1}\mathrm{i}\boldsymbol{\Gamma}_{2l}\mathrm{e}^{\frac{(L-P)\gamma_{l-1}}{4}\boldsymbol{\Gamma}_{2l-1}\mathrm{i}\boldsymbol{\Gamma}_{2l}}\left(\prod_{\substack{n=1\\n\neq l}}^{N}\mathrm{e}^{\frac{L\gamma_{n-1}}{2}\boldsymbol{\Gamma}_{2n-1}\mathrm{i}\boldsymbol{\Gamma}_{2n}}\right), \tag{30}$$

$$\mathbf{V}=\mathrm{e}^{\frac{L\gamma_{l-1}}{4}\boldsymbol{\Gamma}_{2l-1}\mathrm{i}\boldsymbol{\Gamma}_{2l}}\mathrm{e}^{\frac{(L-P)\gamma_{m-1}}{4}\boldsymbol{\Gamma}_{2m-1}\mathrm{i}\boldsymbol{\Gamma}_{2m}}\boldsymbol{\Gamma}_{2l-1}\mathrm{i}\boldsymbol{\Gamma}_{2m}\mathrm{e}^{\frac{L\gamma_{l-1}}{4}\boldsymbol{\Gamma}_{2l-1}\mathrm{i}\boldsymbol{\Gamma}_{2l}}\mathrm{e}^{\frac{(L-P)\gamma_{m-1}}{4}\boldsymbol{\Gamma}_{2m-1}\mathrm{i}\boldsymbol{\Gamma}_{2m}}\left(\prod_{\substack{n=1\\ n\neq l,\, n\neq m}}^{N}\mathrm{e}^{\frac{L\gamma_{n-1}}{2}\boldsymbol{\Gamma}_{2n-1}\mathrm{i}\boldsymbol{\Gamma}_{2n}}\right). \tag{31}$$

(5) For calculating $\mathrm{Tr}\mathbf{U}$ in (29), at first, in terms of (10) and by straightforward calculation we obtain

$$\begin{aligned}
&\mathrm{e}^{\frac{(L-P)\gamma_{l-1}}{4}\boldsymbol{\Gamma}_{2l-1}\mathrm{i}\boldsymbol{\Gamma}_{2l}}\boldsymbol{\Gamma}_{2l-1}\mathrm{i}\boldsymbol{\Gamma}_{2l}\mathrm{e}^{\frac{(L-P)\gamma_{l-1}}{4}\boldsymbol{\Gamma}_{2l-1}\mathrm{i}\boldsymbol{\Gamma}_{2l}}\\
&=\left(\cosh\frac{(L-P)\gamma_{l-1}}{4}+\boldsymbol{\Gamma}_{2l-1}\mathrm{i}\boldsymbol{\Gamma}_{2l}\sinh\frac{(L-P)\gamma_{l-1}}{4}\right)\boldsymbol{\Gamma}_{2l-1}\mathrm{i}\boldsymbol{\Gamma}_{2l}\\
&\quad\times\left(\cosh\frac{(L-P)\gamma_{l-1}}{4}+\boldsymbol{\Gamma}_{2l-1}\mathrm{i}\boldsymbol{\Gamma}_{2l}\sinh\frac{(L-P)\gamma_{l-1}}{4}\right)\\
&=\sinh\frac{(L-P)\gamma_{l-1}}{2}+\boldsymbol{\Gamma}_{2l-1}\mathrm{i}\boldsymbol{\Gamma}_{2l}\cosh\frac{(L-P)\gamma_{l-1}}{2}\\
&=\lim_{\alpha\to 0}\frac{\partial}{\partial\alpha}\left(\cosh\left(\frac{(L-P)\gamma_{l-1}}{2}+\alpha\right)+\boldsymbol{\Gamma}_{2l-1}\mathrm{i}\boldsymbol{\Gamma}_{2l}\sinh\left(\frac{(L-P)\gamma_{l-1}}{2}+\alpha\right)\right)\\
&=\lim_{\alpha\to 0}\frac{\partial}{\partial\alpha}\mathrm{e}^{\left(\frac{(L-P)\gamma_{l-1}}{2}+\alpha\right)\boldsymbol{\Gamma}_{2l-1}\mathrm{i}\boldsymbol{\Gamma}_{2l}},
\end{aligned}$$

and then, substituting the above result into (30), we obtain

$$\mathrm{Tr}\mathbf{U}=\lim_{\alpha\to 0}\frac{\partial}{\partial\alpha}\mathrm{Tr}\left(\mathrm{e}^{\left(\frac{(L-P)\gamma_{l-1}}{2}+\alpha\right)\boldsymbol{\Gamma}_{2l-1}\mathrm{i}\boldsymbol{\Gamma}_{2l}}\left(\prod_{\substack{n=1\\ n\neq l}}^{N}\mathrm{e}^{\frac{L\gamma_{n-1}}{2}\boldsymbol{\Gamma}_{2n-1}\mathrm{i}\boldsymbol{\Gamma}_{2n}}\right)\right). \tag{32}$$

Since (A36) holds for arbitrary values of $\gamma_{n-1}$ $(n=1,2,\cdots,N)$, applying (A36) to (32), we obtain

$$\mathrm{Tr}\boldsymbol{U}=\lim_{\alpha\to 0}\frac{\partial}{\partial\alpha}\left(2\cosh\left(\frac{(L-P)\gamma_{l-1}}{2}+\alpha\right)\right)\prod_{\substack{n=1\\ n\neq l}}^{N}\left(2\cosh\frac{L\gamma_{n-1}}{2}\right)=2\sinh\frac{(L-P)\gamma_{l-1}}{2}\prod_{\substack{n=1\\ n\neq l}}^{N}\left(2\cosh\frac{L\gamma_{n-1}}{2}\right). \tag{33}$$

(6) For calculating $\mathrm{Tr}\mathbf{V}$ in (29), at first, in terms of (10) and (A8), by straightforward calculation we obtain

$$\begin{aligned}
&\mathrm{e}^{L\frac{\gamma_{l-1}}{4}\boldsymbol{\Gamma}_{2l-1}\mathrm{i}\boldsymbol{\Gamma}_{2l}}\mathrm{e}^{L\frac{\gamma_{m-1}}{4}\boldsymbol{\Gamma}_{2m-1}\mathrm{i}\boldsymbol{\Gamma}_{2m}}\boldsymbol{\Gamma}_{2l-1}\mathrm{i}\boldsymbol{\Gamma}_{2m}\mathrm{e}^{\frac{L\gamma_{l-1}}{4}\boldsymbol{\Gamma}_{2l-1}\mathrm{i}\boldsymbol{\Gamma}_{2l}}\mathrm{e}^{\frac{L\gamma_{m-1}}{4}\boldsymbol{\Gamma}_{2m-1}\mathrm{i}\boldsymbol{\Gamma}_{2m}}\\
&=\left(\cosh\frac{L\gamma_{l-1}}{4}+\boldsymbol{\Gamma}_{2l-1}\mathrm{i}\boldsymbol{\Gamma}_{2l}\sinh\frac{L\gamma_{l-1}}{4}\right)\left(\cosh\frac{L\gamma_{m-1}}{4}+\boldsymbol{\Gamma}_{2m-1}\mathrm{i}\boldsymbol{\Gamma}_{2m}\sinh\frac{L\gamma_{m-1}}{4}\right)\boldsymbol{\Gamma}_{2l-1}\mathrm{i}\boldsymbol{\Gamma}_{2m}\\
&\quad\times\left(\cosh\frac{L\gamma_{l-1}}{4}+\boldsymbol{\Gamma}_{2l-1}\mathrm{i}\boldsymbol{\Gamma}_{2l}\sinh\frac{L\gamma_{l-1}}{4}\right)\left(\cosh\frac{L\gamma_{m-1}}{4}+\boldsymbol{\Gamma}_{2m-1}\mathrm{i}\boldsymbol{\Gamma}_{2m}\sinh\frac{L\gamma_{m-1}}{4}\right)\\
&=\boldsymbol{\Gamma}_{2l-1}\mathrm{i}\boldsymbol{\Gamma}_{2m},
\end{aligned}$$

substituting the above result into (31), we obtain

$$\mathrm{Tr}\mathbf{V}=\mathrm{Tr}\left(\boldsymbol{\Gamma}_{2l-1}\mathrm{i}\boldsymbol{\Gamma}_{2m}\left(\prod_{\substack{n=1\\ n\neq l,\, n\neq m}}^{N}\mathrm{e}^{\frac{L\gamma_{n-1}}{2}\boldsymbol{\Gamma}_{2n-1}\mathrm{i}\boldsymbol{\Gamma}_{2n}}\right)\right). \tag{34}$$

However, if we want to use (A36) to calculate (34), then we must write $\boldsymbol{\Gamma}_{2l-1}\mathrm{i}\boldsymbol{\Gamma}_{2m}$ in the form $\mathrm{e}^{\alpha\boldsymbol{\Gamma}_{2l-1}\mathrm{i}\boldsymbol{\Gamma}_{2l}}\mathrm{e}^{\beta\boldsymbol{\Gamma}_{2m-1}\mathrm{i}\boldsymbol{\Gamma}_{2m}}$. For this purpose, by some attempts we find that if we introduce a matrix

$$\mathbf{S}_{lm}=\frac{1}{4}\big(\mathbf{I}+\mathrm{i}\boldsymbol{\Gamma}_{2l-1}-\mathrm{i}\boldsymbol{\Gamma}_{2l}+\boldsymbol{\Gamma}_{2m-1}-\boldsymbol{\Gamma}_{2m}$$
$$+\boldsymbol{\Gamma}_{2l-1}\boldsymbol{\Gamma}_{2l}-\mathrm{i}\boldsymbol{\Gamma}_{2l-1}\boldsymbol{\Gamma}_{2m-1}+\mathrm{i}\boldsymbol{\Gamma}_{2l-1}\boldsymbol{\Gamma}_{2m}-\mathrm{i}\boldsymbol{\Gamma}_{2l}\boldsymbol{\Gamma}_{2m-1}+\mathrm{i}\boldsymbol{\Gamma}_{2l}\boldsymbol{\Gamma}_{2m}+\boldsymbol{\Gamma}_{2m-1}\boldsymbol{\Gamma}_{2m} \tag{35}$$
$$-\boldsymbol{\Gamma}_{2l-1}\boldsymbol{\Gamma}_{2l}\boldsymbol{\Gamma}_{2m-1}+\boldsymbol{\Gamma}_{2l-1}\boldsymbol{\Gamma}_{2l}\boldsymbol{\Gamma}_{2m}+\mathrm{i}\boldsymbol{\Gamma}_{2l-1}\boldsymbol{\Gamma}_{2m-1}\boldsymbol{\Gamma}_{2m}-\mathrm{i}\boldsymbol{\Gamma}_{2l}\boldsymbol{\Gamma}_{2m-1}\boldsymbol{\Gamma}_{2m}+\boldsymbol{\Gamma}_{2l-1}\boldsymbol{\Gamma}_{2l}\boldsymbol{\Gamma}_{2m-1}\boldsymbol{\Gamma}_{2m}\big),$$

then by straightforward calculation we obtain the inverse matrix of $\mathbf{S}_{lm}$ :

$$(\mathbf{S}_{lm})^{-1}=\frac{1}{4}\big(\mathbf{I}-\mathrm{i}\boldsymbol{\Gamma}_{2l-1}+\mathrm{i}\boldsymbol{\Gamma}_{2l}+\boldsymbol{\Gamma}_{2m-1}-\boldsymbol{\Gamma}_{2m}$$
$$-\boldsymbol{\Gamma}_{2l-1}\boldsymbol{\Gamma}_{2l}-\mathrm{i}\boldsymbol{\Gamma}_{2l-1}\boldsymbol{\Gamma}_{2m-1}+\mathrm{i}\boldsymbol{\Gamma}_{2l-1}\boldsymbol{\Gamma}_{2m}-\mathrm{i}\boldsymbol{\Gamma}_{2l}\boldsymbol{\Gamma}_{2m-1}+\mathrm{i}\boldsymbol{\Gamma}_{2l}\boldsymbol{\Gamma}_{2m}-\boldsymbol{\Gamma}_{2m-1}\boldsymbol{\Gamma}_{2m}$$
$$+\boldsymbol{\Gamma}_{2l-1}\boldsymbol{\Gamma}_{2l}\boldsymbol{\Gamma}_{2m-1}-\boldsymbol{\Gamma}_{2l-1}\boldsymbol{\Gamma}_{2l}\boldsymbol{\Gamma}_{2m}+\mathrm{i}\boldsymbol{\Gamma}_{2l-1}\boldsymbol{\Gamma}_{2m-1}\boldsymbol{\Gamma}_{2m}-\mathrm{i}\boldsymbol{\Gamma}_{2l}\boldsymbol{\Gamma}_{2m-1}\boldsymbol{\Gamma}_{2m}+\boldsymbol{\Gamma}_{2l-1}\boldsymbol{\Gamma}_{2l}\boldsymbol{\Gamma}_{2m-1}\boldsymbol{\Gamma}_{2m}\big),$$

and,

$$\boldsymbol{\Gamma}_{2l-1}\mathrm{i}\boldsymbol{\Gamma}_{2m}=-(\mathbf{S}_{lm})^{-1}\boldsymbol{\Gamma}_{2l-1}\mathrm{i}\boldsymbol{\Gamma}_{2l}\boldsymbol{\Gamma}_{2m-1}\mathrm{i}\boldsymbol{\Gamma}_{2m}\mathbf{S}_{lm}\,. \tag{36}$$

On the other hand, according to (10) we have

$$\mathrm{e}^{\frac{\pi\mathrm{i}}{2}\boldsymbol{\Gamma}_{2n-1}\mathrm{i}\boldsymbol{\Gamma}_{2n}}=\cosh\frac{\pi\mathrm{i}}{2}+\boldsymbol{\Gamma}_{2n-1}\mathrm{i}\boldsymbol{\Gamma}_{2n}\sinh\frac{\pi\mathrm{i}}{2}=\cos\frac{\pi}{2}+\boldsymbol{\Gamma}_{2n-1}\mathrm{i}\boldsymbol{\Gamma}_{2n}\mathrm{i}\sin\frac{\pi}{2}=\mathrm{i}\boldsymbol{\Gamma}_{2n-1}\mathrm{i}\boldsymbol{\Gamma}_{2n}\,. \tag{37}$$

According to (36) and (37) we obtain

$$\boldsymbol{\Gamma}_{2l-1}\mathrm{i}\boldsymbol{\Gamma}_{2m}=-(\mathbf{S}_{lm})^{-1}\left(-\mathrm{i}\mathrm{e}^{\frac{\pi\mathrm{i}}{2}\boldsymbol{\Gamma}_{2l-1}\mathrm{i}\boldsymbol{\Gamma}_{2l}}\right)\left(-\mathrm{i}\mathrm{e}^{\frac{\pi\mathrm{i}}{2}\boldsymbol{\Gamma}_{2m-1}\mathrm{i}\boldsymbol{\Gamma}_{2m}}\right)\mathbf{S}_{lm}=(\mathbf{S}_{lm})^{-1}\mathrm{e}^{\frac{\pi\mathrm{i}}{2}\boldsymbol{\Gamma}_{2l-1}\mathrm{i}\boldsymbol{\Gamma}_{2l}}\mathrm{e}^{\frac{\pi\mathrm{i}}{2}\boldsymbol{\Gamma}_{2m-1}\mathrm{i}\boldsymbol{\Gamma}_{2m}}\mathbf{S}_{lm}\,,$$

substituting the above result into (34), we obtain

$$\mathrm{Tr}\mathbf{V}=\mathrm{Tr}\left((\mathbf{S}_{lm})^{-1}\mathrm{e}^{\frac{\pi\mathrm{i}}{2}\boldsymbol{\Gamma}_{2l-1}\mathrm{i}\boldsymbol{\Gamma}_{2l}}\mathrm{e}^{\frac{\pi\mathrm{i}}{2}\boldsymbol{\Gamma}_{2m-1}\mathrm{i}\boldsymbol{\Gamma}_{2m}}\mathbf{S}_{lm}\left(\prod_{\substack{n=1\\ n\neq l,\,n\neq m}}^{N}\mathrm{e}^{\frac{L\gamma_{n-1}}{2}\boldsymbol{\Gamma}_{2n-1}\mathrm{i}\boldsymbol{\Gamma}_{2n}}\right)\right). \tag{38}$$

As can be seen from (35), the matrix $\mathbf{S}_{lm}$ is only related to $\boldsymbol{\Gamma}_{2l-1}$, $\boldsymbol{\Gamma}_{2l}$, $\boldsymbol{\Gamma}_{2m-1}$ and $\boldsymbol{\Gamma}_{2m}$, according to $[\boldsymbol{\Gamma}_n,\boldsymbol{\Gamma}_{2n'-1}\mathrm{i}\boldsymbol{\Gamma}_{2n'}]=0\quad(n\neq n')$ obtained by (A8) we have

$$\left[\mathbf{S}_{lm}\,,\ \prod_{\substack{n=1\\ n\neq l,\,n\neq m}}^{N}\mathrm{e}^{\frac{L\gamma_{n-1}}{2}\boldsymbol{\Gamma}_{2n-1}\mathrm{i}\boldsymbol{\Gamma}_{2n}}\right]=0\,,$$

using the above result and the property $\mathrm{Tr}(\mathbf{AB})=\mathrm{Tr}(\mathbf{BA})$, (38) can be written in the form

$$\begin{aligned}\mathrm{Tr}\mathbf{V}&=\mathrm{Tr}\left((\mathbf{S}_{lm})^{-1}\mathrm{e}^{\frac{\pi\mathrm{i}}{2}\boldsymbol{\Gamma}_{2l-1}\mathrm{i}\boldsymbol{\Gamma}_{2l}}\mathrm{e}^{\frac{\pi\mathrm{i}}{2}\boldsymbol{\Gamma}_{2m-1}\mathrm{i}\boldsymbol{\Gamma}_{2m}}\left(\prod_{\substack{n=1\\ n\neq l,\,n\neq m}}^{N}\mathrm{e}^{\frac{L\gamma_{n-1}}{2}\boldsymbol{\Gamma}_{2n-1}\mathrm{i}\boldsymbol{\Gamma}_{2n}}\right)\mathbf{S}_{lm}\right)\\&=\mathrm{Tr}\left(\mathbf{S}_{lm}(\mathbf{S}_{lm})^{-1}\mathrm{e}^{\frac{\pi\mathrm{i}}{2}\boldsymbol{\Gamma}_{2l-1}\mathrm{i}\boldsymbol{\Gamma}_{2l}}\mathrm{e}^{\frac{\pi\mathrm{i}}{2}\boldsymbol{\Gamma}_{2m-1}\mathrm{i}\boldsymbol{\Gamma}_{2m}}\left(\prod_{\substack{n=1\\ n\neq l,\,n\neq m}}^{N}\mathrm{e}^{\frac{L\gamma_{n-1}}{2}\boldsymbol{\Gamma}_{2n-1}\mathrm{i}\boldsymbol{\Gamma}_{2n}}\right)\right)\\&=\mathrm{Tr}\left(\mathrm{e}^{\frac{\pi\mathrm{i}}{2}\boldsymbol{\Gamma}_{2l-1}\mathrm{i}\boldsymbol{\Gamma}_{2l}}\mathrm{e}^{\frac{\pi\mathrm{i}}{2}\boldsymbol{\Gamma}_{2m-1}\mathrm{i}\boldsymbol{\Gamma}_{2m}}\left(\prod_{\substack{n=1\\ n\neq l,\,n\neq m}}^{N}\mathrm{e}^{\frac{L\gamma_{n-1}}{2}\boldsymbol{\Gamma}_{2n-1}\mathrm{i}\boldsymbol{\Gamma}_{2n}}\right)\right).\end{aligned} \tag{39}$$

For the form given in (39) we can use (A36) straightforwardly, since (A36) holds for arbitrary values of $\gamma_{n-1}$ $(n=1,2,\cdots,N)$; according to (A36) and $\cosh\frac{\pi \mathrm{i}}{2}=\cos\frac{\pi}{2}=0$ we obtain

$$\mathrm{Tr}\mathbf{V}=\left(2\cosh\frac{\pi \mathrm{i}}{2}\right)\left(2\cosh\frac{\pi \mathrm{i}}{2}\right)\left(\prod_{\substack{n=1\\ n\neq l,\, n\neq m}}^{N} 2\cosh\frac{L\gamma_{n-1}}{2}\right)=4\cos^2\frac{\pi}{2}\left(\prod_{\substack{n=1\\ n\neq l,\, n\neq m}}^{N} 2\cosh\frac{L\gamma_{n-1}}{2}\right)=0\,. \tag{40}$$

(7) Substituting the two results given by (33)和(40) into (29), we obtain

$$\begin{aligned}&\mathrm{Tr}\left(\left[\boldsymbol{\Gamma}_1,\left[\boldsymbol{\Gamma}_1,\left(\mathbf{V}_2^{1/2}\mathbf{V}_1\mathbf{V}_2^{1/2}\right)^P\right]\right]\left(\mathbf{V}_2^{1/2}\mathbf{V}_1\mathbf{V}_2^{1/2}\right)^{L-P}\right)\\ &=8\sum_{l=1}^{N}u_{1l}^2\sinh\frac{P\gamma_{l-1}}{2}\sinh\frac{(L-P)\gamma_{l-1}}{2}\prod_{\substack{n=1\\ n\neq l}}^{N}\left(2\cosh\frac{L\gamma_{n-1}}{2}\right)\,,\end{aligned} \tag{41}$$

the term

$$\begin{aligned}\sinh\frac{P\gamma_{l-1}}{2}\sinh\frac{(L-P)\gamma_{l-1}}{2}&=\frac{\cosh\left(\frac{P\gamma_{l-1}}{2}+\frac{(L-P)\gamma_{l-1}}{2}\right)-\cosh\left(\frac{P\gamma_{l-1}}{2}-\frac{(L-P)\gamma_{l-1}}{2}\right)}{2}\\ &=\frac{1}{2}\cosh\frac{L\gamma_{l-1}}{2}-\frac{1}{2}\cosh\left(\left(\frac{L}{2}-P\right)\gamma_{l-1}\right)\,,\end{aligned}$$

substituting this result into (41), we obtain

$$\begin{aligned}&\mathrm{Tr}\left(\left[\boldsymbol{\Gamma}_1,\left[\boldsymbol{\Gamma}_1,\left(\mathbf{V}_2^{1/2}\mathbf{V}_1\mathbf{V}_2^{1/2}\right)^P\right]\right]\left(\mathbf{V}_2^{1/2}\mathbf{V}_1\mathbf{V}_2^{1/2}\right)^{L-P}\right)\\ &=4\sum_{l=1}^{N}u_{1l}^2\cosh\frac{L\gamma_{l-1}}{2}\prod_{\substack{n=1\\ n\neq l}}^{N}\left(2\cosh\frac{L\gamma_{n-1}}{2}\right)-4\sum_{l=1}^{N}u_{1l}^2\cosh\left(\left(\frac{L}{2}-P\right)\gamma_{l-1}\right)\prod_{\substack{n=1\\ n\neq l}}^{N}\left(2\cosh\frac{L\gamma_{n-1}}{2}\right)\\ &=2\sum_{l=1}^{N}u_{1l}^2\prod_{n=1}^{N}\left(2\cosh\frac{L\gamma_{n-1}}{2}\right)-2\sum_{l=1}^{N}u_{1l}^2\frac{\cosh\left(\left(\frac{L}{2}-P\right)\gamma_{l-1}\right)}{\cosh\frac{L\gamma_{l-1}}{2}}\prod_{n=1}^{N}\left(2\cosh\frac{L\gamma_{n-1}}{2}\right)\\ &=2Z_0\left(\sum_{l=1}^{N}u_{1l}^2-\sum_{l=1}^{N}u_{1l}^2\frac{\cosh\left(\left(\frac{L}{2}-P\right)\gamma_{l-1}\right)}{\cosh\frac{L\gamma_{l-1}}{2}}\right)\,,\end{aligned} \tag{42}$$

in the last step of the calculation process in the above formula, we have used the expression of $Z_0$ given by (A2).

According to the first relation in (A29) we have $\sum_{l=1}^{N}u_{1l}^2=1$, and, from the first relation in (A24) we obtain $u_{1m}^2=2\varOmega_{m-1}^2$, where the expression of $\varOmega_{m-1}$ is given by (A25). In terms of these two results, (42) can be written in the form

$$\mathrm{Tr}\left(\left[\boldsymbol{\Gamma}_1,\left[\boldsymbol{\Gamma}_1,\left(\mathbf{V}_2^{1/2}\mathbf{V}_1\mathbf{V}_2^{1/2}\right)^P\right]\right]\left(\mathbf{V}_2^{1/2}\mathbf{V}_1\mathbf{V}_2^{1/2}\right)^{L-P}\right)=2Z_0\left(1-2\sum_{l=1}^{N}\varOmega_{l-1}^2\frac{\cosh\left(\left(\frac{L}{2}-P\right)\gamma_{l-1}\right)}{\cosh\frac{L\gamma_{l-1}}{2}}\right)\,. \tag{43}$$

## 4 Exact expressions and some properties of $\langle \sigma_{1,1}\sigma_{1+P,1}\rangle$

Substituting the result given in (43) into (7), we obtain exact expressions of $\langle \sigma_{1,1}\sigma_{1+P,1}\rangle$:

$$\langle \sigma_{1,1}\sigma_{1+P,1}\rangle = 2\sum_{n=1}^{N} \Omega_{n-1}^2 \frac{\cosh\left[\left(\frac{L}{2}-P\right)\gamma_{n-1}\right]}{\cosh\frac{L\gamma_{n-1}}{2}} = 2\sum_{n=1}^{N} \Omega_{n-1}^2 \frac{e^{-P\gamma_{n-1}}+e^{-(L-P)\gamma_{n-1}}}{1+e^{-L\gamma_{n-1}}}, \tag{44}$$

where $\gamma_{n-1}$ and $\Omega_{n-1}$ are given by (A11) and (A25), respectively.

Some properties of $\langle \sigma_{1,1}\sigma_{1+P,1}\rangle$ given by (44) are as follows.

(1) It is obvious that $\langle \sigma_{1,1}\sigma_{1+P,1}\rangle$ given by (44) satisfy (3).

(2) For $T \ge T_c$, according to (A15), (A16b) and (A21), all $N$ roots of (A12) have forms $x_{n-1} = \cos\left(\frac{n-1}{N}\pi + \frac{\theta_{n-1}}{N}\right)$ $(n=1,2,\cdots,N)$. As $N\to\infty$, according to (A17), (A21) and (A22), $x_{n-1}\to\cos(\omega\pi)$ $(0\le\omega\le 1)$ and $\gamma_{n-1}\to\gamma_{n-1}^{(0)}$; according to (A18) and (A22),

$$\cosh\gamma_{n-1} \to \cosh\gamma(\omega\pi) = \cosh 2K'\cosh 2K - \sinh 2K'\sinh 2K\cos(\omega\pi). \tag{45}$$

According to (A26), as $N\to\infty$,

$$\Omega_{n-1}^2 \to \frac{\sinh^2 2K'\cosh^2 K}{N}\frac{\sin^2(\omega\pi)}{\sinh^2\gamma(\omega\pi)} \qquad (n=1,2,\cdots,N). \tag{46}$$

Combining the above results, as $N\to\infty$, (44) becomes

$$\begin{aligned}\lim_{N\to\infty}\langle \sigma_{1,1}\sigma_{1+P,1}\rangle &= 2\sinh^2 2K'\cosh^2 K\int_0^1 d\omega \frac{\sin^2(\omega\pi)}{\sinh^2\gamma(\omega\pi)}\frac{\cosh\left[\left(\frac{L}{2}-P\right)\gamma(\omega\pi)\right]}{\cosh\frac{L\gamma(\omega\pi)}{2}} \\ &= 2\sinh^2 2K'\cosh^2 K\int_0^\pi \frac{d\omega}{\pi}\frac{\sin^2\omega}{\sinh^2\gamma(\omega)}\frac{e^{-P\gamma(\omega)}+e^{-(L-P)\gamma(\omega)}}{1+e^{-L\gamma(\omega)}}.\end{aligned} \tag{47}$$

where the function $\gamma(\omega)$ is determined by

$$\cosh\gamma(\omega) = \cosh 2K'\cosh 2K - \sinh 2K'\sinh 2K\cos\omega, \qquad \gamma(\omega)>0. \tag{48}$$

And, further, for (47), as $L\to\infty$, we obtain

$$\lim_{L\to\infty}\lim_{N\to\infty}\langle \sigma_{1,1}\sigma_{1+P,1}\rangle = \begin{cases}\Phi(Q_1), & P=Q_1 \text{ or } P=L-Q_1; \\ 0, & P=\left[\frac{L}{2}\right]\pm Q_2,\end{cases} \tag{49}$$

where

$$\Phi(Q_1) = 2\sinh^2 2K'\cosh^2 K\int_0^\pi \frac{d\omega}{\pi}\frac{\sin^2\omega}{\sinh^2\gamma(\omega)} e^{-Q_1\gamma(\omega)}, \tag{50}$$

$Q_1 = 1,2,3,\cdots$ be small positive integer such that $\lim_{L\to\infty}(L-Q_1)=\infty$, and $Q_2 = 0,1,2,3,\cdots$ be zero or small positive integer such that $\lim_{L\to\infty}\left(\left[\frac{L}{2}\right]\pm Q_2\right)=\infty$.

We conclude in Sec.2 that, for $\left\langle \sigma_{1,1}\sigma_{1+P,1} \right\rangle$, when $P = Q_1$ or $P = L - Q_1$, where $Q_1 = 1, 2, 3, \cdots$ are small positive integers, $\left\langle \sigma_{1,1}\sigma_{1+P,1} \right\rangle$ belong to short range order; and, when $P = \left[\frac{L}{2}\right] \pm Q_2$, where $Q_2 = 0, 1, 2, 3, \cdots$ are zero or small positive integers, $\left\langle \sigma_{1,1}\sigma_{1+P,1} \right\rangle$ belong to long range order. Hence, from (50) and (51) we see that, in the case of $T \ge T_c$ and in the thermodynamic limit, short range order do not vanish but long range order vanish; this conclusion is in accord with that in Introduction of this paper.

The order of taking limits of $L$ and $N$ in (49) is $\lim_{L\to\infty}\lim_{N\to\infty}\left\langle \sigma_{1,1}\sigma_{1+P,1} \right\rangle$, conversely, if the order of taking limits of $L$ and $N$ is $\lim_{N\to\infty}\lim_{L\to\infty}\left\langle \sigma_{1,1}\sigma_{1+P,1} \right\rangle$, then as $L \to \infty$, (44) becomes

$$\lim_{L\to\infty}\left\langle \sigma_{1,1}\sigma_{1+P,1} \right\rangle = \begin{cases} 2\sum_{n=1}^{N} \Omega_{n-1}^2 \mathrm{e}^{-Q_1\gamma_{n-1}}, & P = Q_1 \text{ or } P = L - Q_1; \\ 0, & P = \left[\frac{L}{2}\right] \pm Q_2; \end{cases}$$

and then, according to (46) we obtain

$$\lim_{N\to\infty}\lim_{L\to\infty}\left\langle \sigma_{1,1}\sigma_{1+P,1} \right\rangle = \begin{cases} \Phi(Q_1), & P = Q_1 \text{ or } P = L - Q_1; \\ 0, & P = \left[\frac{L}{2}\right] \pm Q_2, \end{cases} \tag{51}$$

where $\Phi(Q_1)$ is given by (50).

Comparing (49) with (51), we obtain

$$\lim_{L\to\infty}\lim_{N\to\infty}\left\langle \sigma_{1,1}\sigma_{1+P,1} \right\rangle = \lim_{N\to\infty}\lim_{L\to\infty}\left\langle \sigma_{1,1}\sigma_{1+P,1} \right\rangle, \tag{52}$$

hence, the different orders of taking limits of $L$ and $N$ have no impact on $\left\langle \sigma_{1,1}\sigma_{1+P,1} \right\rangle$.

(3) Certain properties of $\left\langle \sigma_{1,1}\sigma_{1+P,1} \right\rangle$ for $T < T_c$

(i) Expresses of $\left\langle \sigma_{1,1}\sigma_{1+P,1} \right\rangle$

For $T < T_c$, we write (44) in the form

$$\left\langle \sigma_{1,1}\sigma_{1+P,1} \right\rangle = \Lambda_P + Y_P, \tag{53}$$

$$\Lambda_P = 2\Omega_0^2 \frac{\cosh\left[\left(\frac{L}{2} - P\right)\gamma_0\right]}{\cosh\frac{L\gamma_0}{2}}, \tag{54}$$

$$Y_P = 2\sum_{n=2}^{N} \Omega_{n-1}^2 \frac{\cosh\left[\left(\frac{L}{2} - P\right)\gamma_{n-1}\right]}{\cosh\frac{L\gamma_{n-1}}{2}} = 2\sum_{n=2}^{N} \Omega_{n-1}^2 \frac{\mathrm{e}^{-P\gamma_{n-1}} + \mathrm{e}^{-(L-P)\gamma_{n-1}}}{1 + \mathrm{e}^{-L\gamma_{n-1}}}. \tag{55}$$

In the case of $T \le T_c$, according to (A15), we still have $x_{n-1} = \cos\left(\frac{n-1}{N}\pi + \frac{\theta_{n-1}}{N}\right)$

$(n = 2, 3, \cdots, N)$, therefore, as $N \to \infty$, both (45) and (46) still hold for $n = 2, 3, \cdots, N$, and, thus, using the method of proving (52), for $Y_P$ given by (55) we can prove

$$\lim_{N\to\infty}\lim_{L\to\infty} Y_P = \lim_{L\to\infty}\lim_{N\to\infty} Y_P = \begin{cases} \Phi(Q_1), & P = Q_1 \text{ or } P = L - Q_1 ; \\ 0, & P = \left[\dfrac{L}{2}\right] \pm Q_2 , \end{cases} \tag{56}$$

where $\Phi(Q_1)$ is given by (50).

(ii) Certain properties of $\Lambda_P$ given by (54) in the cases of $T < T_c$

Substituting the two expressions of $\gamma_0$ and $\Omega_0^2$ given by (A22) and (A27) respectively in the cases of $T < T_c$ into (54), $\Lambda_P$ can be written in the form

$$\Lambda_P \approx \frac{\cosh 2K - \cosh 2K'}{2\sinh^2 K} \frac{\cosh\left[(L-2P)\dfrac{\cosh 2K - \cosh 2K'}{\sinh 2K}\left(\dfrac{\tanh K'}{\tanh K}\right)^N\right]}{\cosh\left[L\dfrac{\cosh 2K - \cosh 2K'}{\sinh 2K}\left(\dfrac{\tanh K'}{\tanh K}\right)^N\right]} . \tag{57}$$

For this case, since $K' < K$ holds true when $T < T_c$ holds true, hence, $\dfrac{\tanh K'}{\tanh K} < 1$, and, thus, $\left(\dfrac{\tanh K'}{\tanh K}\right)^N$ becomes exponentially little and does vanish rapidly as $N \to \infty$, we therefore have

$$\lim_{N\to\infty}(L-2P)\left(\frac{\tanh K'}{\tanh K}\right)^N = 0 , \qquad \lim_{N\to\infty} L\left(\frac{\tanh K'}{\tanh K}\right)^N = 0$$

even if $L = N^k$ is power of $N$. Hence, in this case, in terms of $\cosh 0 = 1$, $\Lambda_P$ given by (57) has property

$$\lim_{N\to\infty} \Lambda_P = \frac{\cosh 2K - \cosh 2K'}{2\sinh^2 K}, \tag{58}$$

and, further,

$$\lim_{L\to\infty}\lim_{N\to\infty} \Lambda_P = \frac{\cosh 2K - \cosh 2K'}{2\sinh^2 K} . \tag{59}$$

On the other hand, from (57) we have

$$\lim_{L\to\infty} \Lambda_P = \frac{\cosh 2K - \cosh 2K'}{2\sinh^2 K} \times \begin{cases} \exp\left[-2Q_1\dfrac{\cosh 2K - \cosh 2K'}{\sinh 2K}\left(\dfrac{\tanh K'}{\tanh K}\right)^N\right], & P = Q_1 \text{ or } P = L - Q_1 ; \\ 0, & P = \left[\dfrac{L}{2}\right] \pm Q_2 , \end{cases} \tag{60}$$

in the case of $T$ far away from $T_c$, $\left(\dfrac{\tanh K'}{\tanh K}\right)^N \to 0$ exponentially as $N \to \infty$, and, thus, $\lim_{N\to\infty} \exp\left[-2Q_1\dfrac{\cosh 2K - \cosh 2K'}{\sinh 2K}\left(\dfrac{\tanh K'}{\tanh K}\right)^N\right] = \mathrm{e}^0 = 1$, we therefore obtain

$$\lim_{N\to\infty}\lim_{L\to\infty} \Lambda_P = \begin{cases} \dfrac{\cosh 2K - \cosh 2K'}{2\sinh^2 K}, & P = Q_1 \text{ or } P = L - Q_1 ; \\ 0, & P = \left[\dfrac{L}{2}\right] \pm Q_2 . \end{cases} \tag{61}$$

Comparing (59) with (61), we obtain

$$\lim_{L\to\infty}\lim_{N\to\infty} \Lambda_P \neq \lim_{N\to\infty}\lim_{L\to\infty} \Lambda_P , \tag{62}$$

Hence, the different orders of taking limits of $L$ and $N$ for $\Lambda_P$ have an impact.

(iii) The value of $\Lambda_P$ when $T = T_c$

According to definitions of $K'$ and $K$ given by (A4), $\Lambda_P$ in (57) can be written in the form

$$\Lambda_P(T) \approx \frac{\cosh\frac{2J}{kT} - \coth\frac{2J'}{kT}}{2\sinh^2\frac{J}{kT}} \frac{\cosh\left[(L-2P)\dfrac{\cosh\frac{2J}{kT} - \coth\frac{2J'}{kT}}{\sinh\frac{2J}{kT}}\left(\dfrac{e^{-\frac{2J'}{kT}}}{\tanh\frac{J}{kT}}\right)^N\right]}{\cosh\left[L\dfrac{\cosh\frac{2J}{kT} - \coth\frac{2J'}{kT}}{\sinh\frac{2J}{kT}}\left(\dfrac{e^{-\frac{2J'}{kT}}}{\tanh\frac{J}{kT}}\right)^N\right]}, \tag{63}$$

especially, at the critical point, i.e., $T = T_c$ ,

$$\Lambda_P(T_c) = \frac{\cosh\frac{2J}{kT_c} - \coth\frac{2J'}{kT_c}}{2\sinh^2\frac{J}{kT_c}} \frac{\cosh\left[(L-2P)\dfrac{\cosh\frac{2J}{kT_c} - \coth\frac{2J'}{kT_c}}{\sinh\frac{2J}{kT_c}}\left(\dfrac{e^{-\frac{2J'}{kT_c}}}{\tanh\frac{J}{kT_c}}\right)^N\right]}{\cosh\left[L\dfrac{\cosh\frac{2J}{kT_c} - \coth\frac{2J'}{kT_c}}{\sinh\frac{2J}{kT_c}}\left(\dfrac{e^{-\frac{2J'}{kT_c}}}{\tanh\frac{J}{kT_c}}\right)^N\right]}. \tag{64}$$

According to (A19b), as $N \to \infty$ , the critical temperature $T_c$ of the system is determined by

$$\sinh\frac{2J}{kT_c}\cosh\frac{2J'}{kT_c} - \cosh\frac{2J}{kT_c} = 0 ; \tag{65a}$$

this result is in accord with that of the two-dimensional rectangular Ising model with periodic-periodic boundary condition[1~3]; it is easy to prove following equivalent forms of (67a):

$$\tanh\frac{J}{kT_c} = e^{-\frac{2J'}{kT_c}} , \qquad \sinh\frac{2J'}{kT_c}\sinh\frac{2J}{kT_c} = 1 , \qquad \cosh\frac{2J}{kT_c} = \coth\frac{2J'}{kT_c} . \tag{65b}$$

According to the first relation in (65b), $e^{\frac{2J'}{kT_c}}\tanh\frac{J}{kT_c} = 1$ ; hence, notice $K' < K$ holds true when $T < T_c$ holds true, although $\left(e^{\frac{2J'}{kT}}\tanh\frac{J}{kT}\right)^N = \left(\frac{\tanh K}{\tanh K'}\right)^N \to \infty$ exponentially as $N \to \infty$ ,

at the critical point $T = T_c$, $\left( e^{\frac{2J'}{kT_c}} \tanh \frac{J}{kT_c} \right)^N = 1^N = 1$.

Based on the above result, According to the third relation in (65b), from (64) we obtain

$$\Lambda_P(T_c) = 0 . \tag{66}$$

(iv) Certain properties of $\langle \sigma_{1,1}\sigma_{1+P,1} \rangle$ for $T < T_c$

In terms of (53), (56) and (59), we obtain

$$\lim_{L\to\infty}\lim_{N\to\infty}\langle \sigma_{1,1}\sigma_{1+P,1} \rangle = \begin{cases} \dfrac{\cosh 2K - \cosh 2K'}{2\sinh^2 K} + \Phi(Q_1), & P = Q_1 \text{ or } P = L - Q_1 ; \\ \dfrac{\cosh 2K - \cosh 2K'}{2\sinh^2 K}, & P = \left[\dfrac{L}{2}\right] \pm Q_2 , \end{cases} \tag{67}$$

As can be seen by comparing (49) with (67), in the thermodynamic limit and for $P = \left[\frac{L}{2}\right] \pm Q_2$, $Q_2 = 0, 1, 2, 3, \cdots$ be zero or small positive integer, $\lim_{L\to\infty}\lim_{N\to\infty}\langle \sigma_{1,1}\sigma_{[L/2]\pm Q_2+1,1} \rangle$ vanishes for $T > T_c$ but read $\frac{\cosh 2K - \cosh 2K'}{2\sinh^2 K}$ for $T \le T_c$. Hence, as the decrease of the temperature, once the system crosses its critical temperature $T_c$, long range order emerge.

In terms of (53), (56) and (61), we obtain

$$\lim_{N\to\infty}\lim_{L\to\infty}\langle \sigma_{1,1}\sigma_{1+P,1} \rangle = \begin{cases} \dfrac{\cosh 2K - \cosh 2K'}{2\sinh^2 K} + \Phi(Q_1), & P = Q_1 \text{ or } P = L - Q_1 ; \\ 0, & P = \left[\dfrac{L}{2}\right] \pm Q_2 . \end{cases} \tag{68}$$

Comparing (67) with (68), we obtain

$$\lim_{L\to\infty}\lim_{N\to\infty}\langle \sigma_{1,1}\sigma_{1+P,1} \rangle \ne \lim_{N\to\infty}\lim_{L\to\infty}\langle \sigma_{1,1}\sigma_{1+P,1} \rangle, \tag{69}$$

concretely, for $P = Q_1$ or $P = L - Q_1$, where $Q_1 = 1, 2, 3, \cdots$ are small positive integers,

$$\lim_{L\to\infty}\lim_{N\to\infty}\langle \sigma_{1,1}\sigma_{1+P,1} \rangle = \lim_{N\to\infty}\lim_{L\to\infty}\langle \sigma_{1,1}\sigma_{1+P,1} \rangle ;$$

this result is the same as (52); however, for $P = \left[\frac{L}{2}\right] \pm Q_2$, where $Q_2 = 0, 1, 2, 3, \cdots$ are zero or small positive integers, $\lim_{N\to\infty}\lim_{L\to\infty}\langle \sigma_{1,1}\sigma_{1+P,1} \rangle$ given by (68) is missing key item $\frac{\cosh 2K - \cosh 2K'}{2\sinh^2 K}$ compared to $\lim_{L\to\infty}\lim_{N\to\infty}\langle \sigma_{1,1}\sigma_{1+P,1} \rangle$ given by (67). Hence, the different orders of taking limits of *L* and *N* for $\langle \sigma_{1,1}\sigma_{1+P,1} \rangle$ have an impact on the results for $T < T_c$, this property is different from that given by (52) for $T \ge T_c$.

We conclude in Sec.2 that, for $\langle \sigma_{1,1}\sigma_{1+P,1} \rangle$, when $P = Q_1$ or $P = L - Q_1$, where $Q_1 = 1, 2, 3, \cdots$ are small positive integers, $\langle \sigma_{1,1}\sigma_{1+P,1} \rangle$ belong to short range order; and, when $P=\left[\frac{L}{2}\right]\pm Q_2$, where $Q_2=0,1,2,3,\cdots$ are zero or small positive integers, $\langle\sigma_{1,1}\sigma_{1+P,1}\rangle$ belong to long range order. Hence, as can be seen from (68), in the case of $T<T_{\mathrm{c}}$ and in the thermodynamic limit, if the order of taking limits of $L$ and $N$ is $\lim_{N\to\infty}\lim_{L\to\infty}\langle\sigma_{1,1}\sigma_{1+P,1}\rangle$, then short range order do not vanish but long range order vanish.

From the above discussion, it is concluded that if the order of limits between $L$ and $N$ is $\lim_{N\to\infty}\lim_{L\to\infty}\langle\sigma_{1,1}\sigma_{1+P,1}\rangle$, then such order of limits cannot reveal long range order. On the other hand, from the numerical calculation results of $\langle\sigma_{1,1}\sigma_{1+P,1}\rangle$ with finite sizes in Sec.5 and Fig.3 ~ Fig.5, it can be seen that long range order exist, hence, $\lim_{L\to\infty}\lim_{N\to\infty}\langle\sigma_{1,1}\sigma_{1+P,1}\rangle$ can correctly describe properties of $\langle\sigma_{1,1}\sigma_{1+P,1}\rangle$, but $\lim_{N\to\infty}\lim_{L\to\infty}\langle\sigma_{1,1}\sigma_{1+P,1}\rangle$ cannot correctly describe properties of $\langle\sigma_{1,1}\sigma_{1+P,1}\rangle$.

The above results indicate that for the thermodynamic limit, the correct order of limits is very important.

(4) Expressions given by (49) and (67) are that of $\langle\sigma_{1,1}\sigma_{1+P,1}\rangle$ in the thermodynamic limit under the limit order $\lim_{L\to\infty}\lim_{N\to\infty}\cdots$, in Appendix B of this paper, we prove that (49) and (67) are the same as that given by Ref.[1].

However, according to (68) we conclude that $\lim_{N\to\infty}\lim_{L\to\infty}\langle\sigma_{1,1}\sigma_{1+P,1}\rangle$ are different from that expressions of $\langle\sigma_{1,1}\sigma_{1+P,1}\rangle$ in the thermodynamic limit given by Ref.[1] for $T<T_{\mathrm{c}}$ and $P=\left[\frac{L}{2}\right]\pm Q_2$.

## 3 Demonstration of numerical calculation results of $\langle\sigma_{1,1}\sigma_{1+P,1}\rangle$

In this section, we demonstrate some numerical calculation results of $\langle\sigma_{1,1}\sigma_{1+P,1}\rangle$ through diagrams.

For this purpose, we first investigate the critical temperature $T_{\mathrm{c}}$ in terms of (A19b), for simplicity, we set $J'=J$, (A19b) thus becomes

$$\cosh\frac{2J}{kT_{\mathrm{c}}}\left(\sinh\frac{2J}{kT_{\mathrm{c}}}-1\right)=\frac{1}{N}. \tag{70}$$

As can be seen from (70), $T_{\mathrm{c}}$ is relevant to $N$, and, it is easy to prove that when $N=1$ and $N\to\infty$, $T_{\mathrm{c}}$ reaches its minimum value and maximum value, that is, $x_{\mathrm{c}}=\frac{J}{kT_{\mathrm{c}}}$ reaches its maximum value and minimum value, respectively. Substituting $N=1$ and $N\to\infty$ into (70), we obtain concrete numerical values of minimum value and maximum value of $x_c$ :

$$x_{c(\min)} = 0.4406867935097718 \quad \text{for } N\to\infty , \qquad x_{c(\max)} = 0.6093778634360064 \quad \text{for } N=1 . \tag{71}$$

Substituting $K'$ and $K$ defined by (A4) into (A11) and (A25), for the case $J'=J$ , we obtain

$$\cosh\gamma_{n-1} = \cosh\frac{2J}{kT}\coth\frac{2J}{kT} - x_{n-1} \qquad (\gamma_{n-1}>0) , \tag{72}$$

$$\Omega^2_{n-1} = \frac{1}{4\sinh^2\frac{J}{kT}} \frac{1-x^2_{n-1}}{N\sinh^2\gamma_{n-1} + \cosh\gamma_{n-1}\coth\frac{2J}{kT} - \cosh\frac{2J}{kT}} . \tag{73}$$

In terms of (70), (72), (73) and (44), we can make numerical calculation for $\langle\sigma_{1,1}\sigma_{1+P,1}\rangle$ and plot based on the numerical calculation results.

For illustrations of $\langle\sigma_{1,1}\sigma_{1+P,1}\rangle$ , there is only a very slight difference among $L>100$ , $N>20$ and $L=100$ , $N=20$ . Fig.3 is as an example, in which $x=\frac{J}{kT}$ and $L=100$ , hence, $\langle\sigma_{1,1}\sigma_{50,1}\rangle$ belongs to long range order. As can be seen from Fig.3, for illustrations of $\langle\sigma_{1,1}\sigma_{50,1}\rangle$, there is only a very slight difference among $N=20$ , 40 and 60, and, the curve for $N=60$ completely overlaps with the curve for $N=40$ .

As can be seen from Fig.3, with the increase in temperature, that is, the decrease in $x$ , once the temperature crosses the critical point, correlation functions vanish; moreover, this characteristic becomes very pronounced well before $L\to\infty$ and $N\to\infty$ .

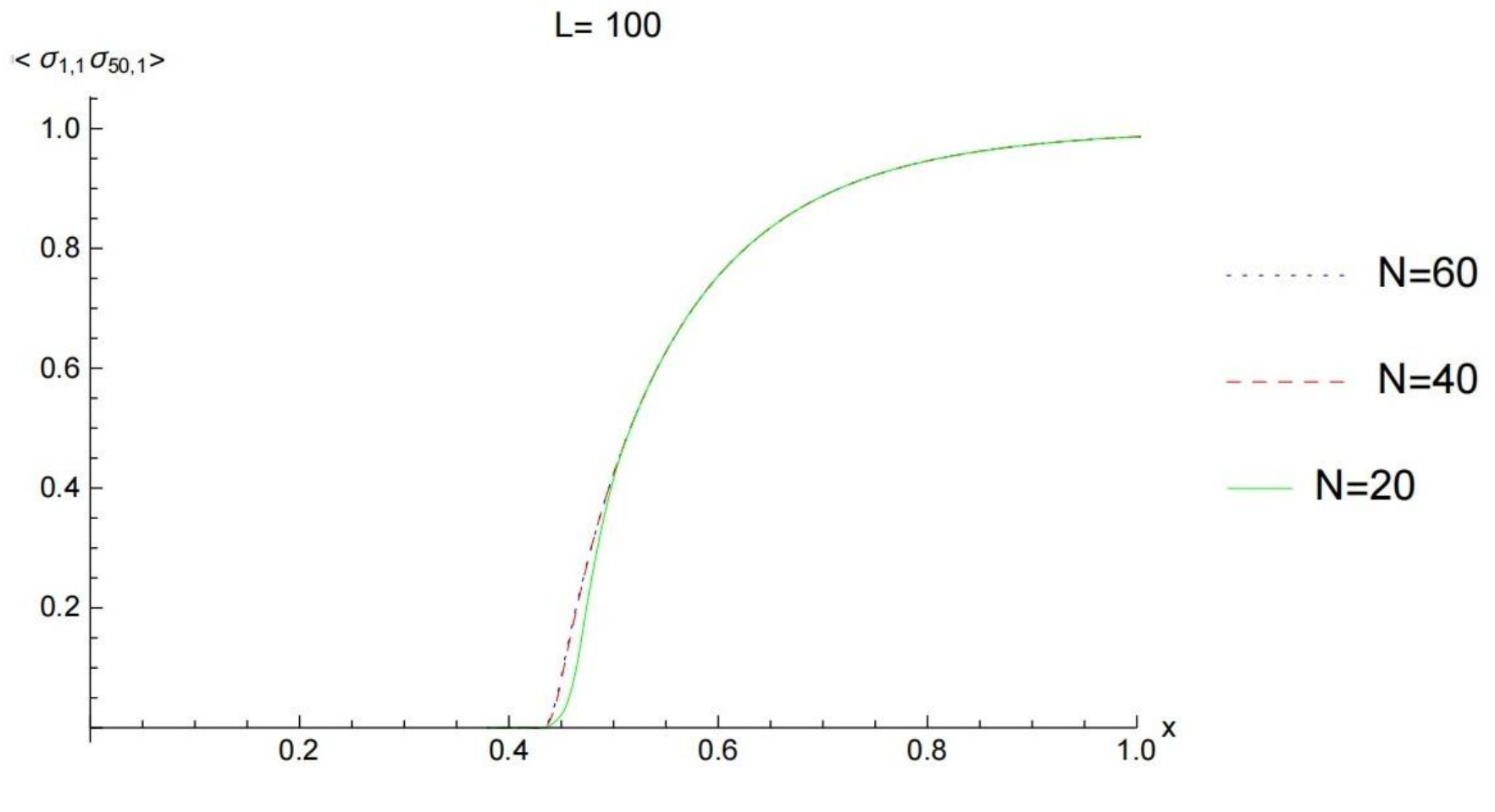

Fig.3 Comparison of increasing of $N$

Due to the above property, for illustrations of $\langle\sigma_{1,1}\sigma_{1+P,1}\rangle$ , we only consider the situations of $L\le 100$ and $N\le 20$ , some examples are shown in Fig.4 and Fig.5.

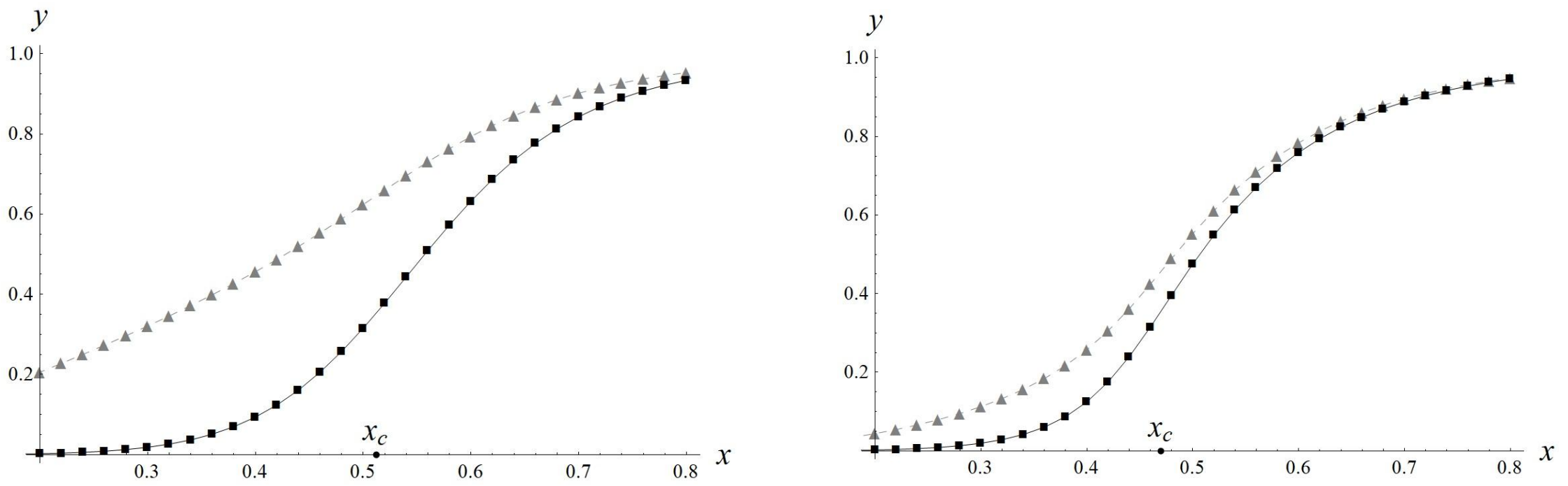

(a) $N=3$ (b) $N=8$

▲: $\langle\sigma_{1,1}\sigma_{2,1}\rangle$; ■: $\langle\sigma_{1,1}\sigma_{5,1}\rangle$ ▲: $\langle\sigma_{1,1}\sigma_{3,1}\rangle$; ■: $\langle\sigma_{1,1}\sigma_{5,1}\rangle$

Fig.4 Some numerical calculation results of $\langle\sigma_{1,1}\sigma_{1+P,1}\rangle$: $L=10$

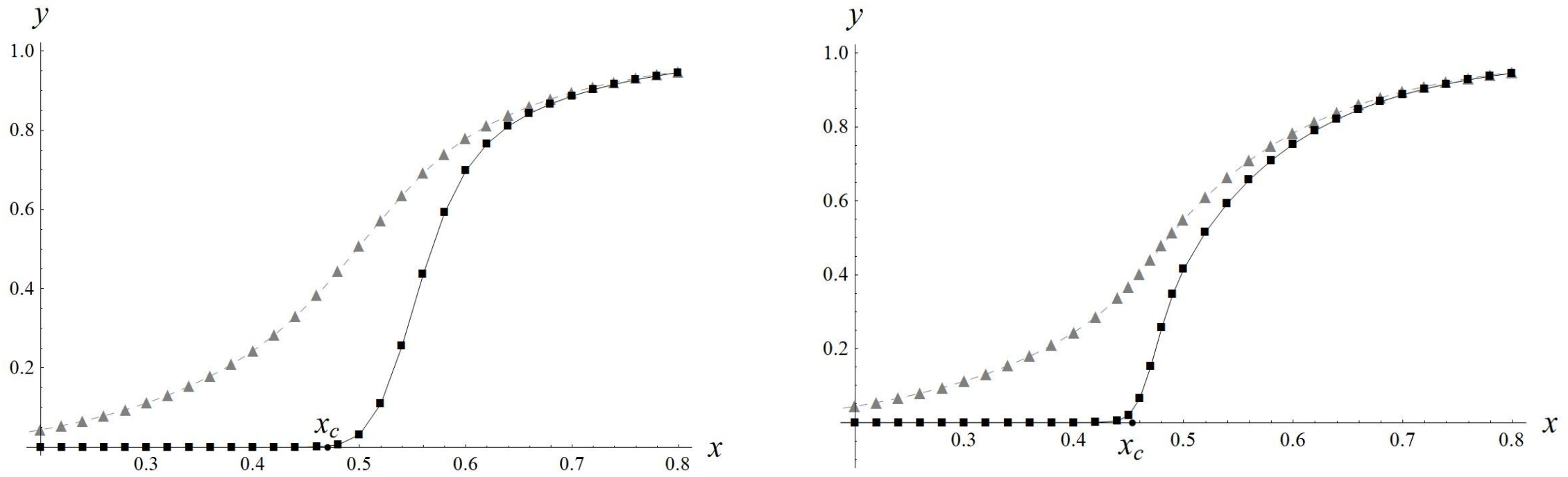

(a) $N=8$ (b) $N=20$

▲: $\langle\sigma_{1,1}\sigma_{3,1}\rangle$; ■: $\langle\sigma_{1,1}\sigma_{50,1}\rangle$ ▲: $\langle\sigma_{1,1}\sigma_{3,1}\rangle$; ■: $\langle\sigma_{1,1}\sigma_{50,1}\rangle$

Fig.5 Some numerical calculation results of $\langle\sigma_{1,1}\sigma_{1+P,1}\rangle$: $L=100$

Tab.1 Some special values of $\langle\sigma_{1,1}\sigma_{1+P,1}\rangle$ when $x=\dfrac{J}{kT}$ near $x_c=\dfrac{J}{kT_c}$

| | | $x=0.2$ | $x=0.4$ |
|---|---|---|---|
| $L=100$, $N=8$ | $\langle\sigma_{1,1}\sigma_{3,1}\rangle$ | 0.0439737234698188 | 0.2437836075425074 |
| | $\langle\sigma_{1,1}\sigma_{50,1}\rangle$ | $1.0553650659827587\times10^{-28}$ | $1.2956901571326502\times10^{-7}$ |
| $L=100$, $N=20$ | $\langle\sigma_{1,1}\sigma_{3,1}\rangle$ | 0.0439737235127527 | 0.2443536642839756 |
| | $\langle\sigma_{1,1}\sigma_{50,1}\rangle$ | $1.1486240314290135\times10^{-28}$ | $2.733658098544861\times10^{-6}$ |

In Fig.4 and Fig.5, $x=\dfrac{J}{kT}$ and $y=\langle\sigma_{1,1}\sigma_{1+P,1}\rangle$, $\langle\sigma_{1,1}\sigma_{1+P,1}\rangle$ illustrated by dotted lines and marked by ▲ belong to short range order, $\left\langle \sigma_{1,1}\sigma_{1+P,1} \right\rangle$ illustrated by full lines and marked by ■ belong to long range order. For highlighting the emergence behavior of $\left\langle \sigma_{1,1}\sigma_{1+P,1} \right\rangle$ when $T$ near the critical temperature $T_c$, we present some special values of $\left\langle \sigma_{1,1}\sigma_{1+P,1} \right\rangle$ when $x = \frac{J}{kT}$ near $x_c = \frac{J}{kT_c}$ in Tab.1.

As can be seen from Fig.4, Fig.5 and Tab.1, $\left\langle \sigma_{1,1}\sigma_{1+P,1} \right\rangle$ have following properties.

(1) When $L$ is a small number, for example, $L = 10$ shown in Fig.4, both $\left\langle \sigma_{1,1}\sigma_{2,1} \right\rangle$ and $\left\langle \sigma_{1,1}\sigma_{3,1} \right\rangle$ marked by ▲ in Fig.4(a) and Fig.4(b) belong to short range order, both $\left\langle \sigma_{1,1}\sigma_{5,1} \right\rangle$ marked by ■ in Fig.4(a) and Fig.4(b) belong to long range order, as can be seen from Fig.4, the critical temperature $T_c$, that is, $x_c \left( = \frac{J}{kT_c} \right)$, has no actual effect on correlation functions.

(2) When $L$ is very large, for example, $L = 100$ shown in Fig.5, from Fig.5 and Tab.1, the following conclusions are drawn.

(i) The critical temperature has virtually no effect on short range order $\left\langle \sigma_{1,1}\sigma_{3,1} \right\rangle$ marked by ▲ in Fig.5(a) and Fig.5(b); as can be seen from Tab.1, when the temperature crosses the critical temperature, that is, $x = \frac{J}{kT}$ near $x_c = \frac{J}{kT_c}$, the difference in values of $\left\langle \sigma_{1,1}\sigma_{3,1} \right\rangle$ between $x = 0.2$ and $x = 0.4$ is not significant.

(ii) The critical temperature has a significant impact on long range order $\left\langle \sigma_{1,1}\sigma_{50,1} \right\rangle$ marked by ■ in Fig.5(a) and Fig.5(b). Actually, graphical method can no longer display change in values of $\left\langle \sigma_{1,1}\sigma_{50,1} \right\rangle$, as can be seen from Tab.1, when the temperature crosses the critical temperature, that is, $x = \frac{J}{kT}$ near $x_c = \frac{J}{kT_c}$, values of $\left\langle \sigma_{1,1}\sigma_{50,1} \right\rangle$ at $x = 0.2$ and $x = 0.4$ respectively are a far cry from each other.

(iii) When $T >> T_c$, that is, $x << x_c$, there is a huge difference between short range order $\left\langle \sigma_{1,1}\sigma_{3,1} \right\rangle$ and long range order $\left\langle \sigma_{1,1}\sigma_{50,1} \right\rangle$, actually, in this case, $\left\langle \sigma_{1,1}\sigma_{50,1} \right\rangle$ is almost zero. On the other hand, when $T << T_c$, that is, $x >> x_c$, there is only a small difference between short range order $\left\langle \sigma_{1,1}\sigma_{3,1} \right\rangle$ and long range order $\left\langle \sigma_{1,1}\sigma_{50,1} \right\rangle$, moreover, as $N$ increases, this difference becomes progressively smaller.

## 6 Summary

In this paper, we present exact expressions of correlation functions $\left\langle\sigma_{1,1}\sigma_{1+P,1}\right\rangle$ between two spins in the boundary row of the two-dimensional rectangular Ising model with periodic-free boundary conditions in absence of magnetic field and finite size.

After obtaining exact expressions of $\left\langle\sigma_{1,1}\sigma_{1+P,1}\right\rangle$, whose some properties are discussed.

It is shown clearly the fundamental property of the Ising model that the long range order emerge as the temperature decrease. Expressions of $\left\langle\sigma_{1,1}\sigma_{1+P,1}\right\rangle$ in the thermodynamic limit varies depending on the order of taking limits of $L$ and $N$, concretely, in the case of the temperature is below the critical temperature, $\lim_{L\to\infty}\lim_{N\to\infty}\left\langle\sigma_{1,1}\sigma_{1+P,1}\right\rangle$ leads to that the long range order emerge, but $\lim_{N\to\infty}\lim_{L\to\infty}\left\langle\sigma_{1,1}\sigma_{1+P,1}\right\rangle$ leads to that the long range order vanish, and, thus, $\lim_{L\to\infty}\lim_{N\to\infty}\left\langle\sigma_{1,1}\sigma_{1+P,1}\right\rangle \neq \lim_{N\to\infty}\lim_{L\to\infty}\left\langle\sigma_{1,1}\sigma_{1+P,1}\right\rangle$. The impact of different sizes on $\left\langle\sigma_{1,1}\sigma_{1+P,1}\right\rangle$ is illustrated with the aid of diagrams. In Appendix B of this paper we prove that $\lim_{L\to\infty}\lim_{N\to\infty}\left\langle\sigma_{1,1}\sigma_{1+P,1}\right\rangle$ are the same as expressions of $\left\langle\sigma_{1,1}\sigma_{1+P,1}\right\rangle$ in the thermodynamic limit in absence of magnetic field given by Ref.[1].

The relationship between the results obtained in this paper and some recent research on the two-dimensional Ising model, e.g., form factor expansions, Painlevé representations, conformal field theory and finite-size corrections, etc[6-9], will be further systematically studied.

## Appendix A The calculation of the partition function given by Eq.(2)

In this Appendix, we list some results needed in this paper given in Refs.[4, 5].

### A.1 The matrix form of the partition function $Z$ given by Eq.(2)

In Ref.[4], the partition function $Z$ given by (2) is expressed as the matrix form:

$$Z=\left(2\sinh\frac{2J'}{kT}\right)^{\frac{LN}{2}}Z_0,. \tag{A1}$$

$$Z_0=\mathrm{Tr}\left(\left(\mathbf{V}_2^{1/2}\mathbf{V}_1\mathbf{V}_2^{1/2}\right)^L\right), \tag{A2}$$

where

$$\mathbf{V}_1=\exp\left(K'\sum_{n=1}^{N}\mathbf{X}_n\right)=\exp\left(K'\sum_{n=1}^{N}\mathbf{\Gamma}_{2n-1}\mathrm{i}\mathbf{\Gamma}_{2n}\right),$$

$$\mathbf{V}_2^{1/2}=\exp\left(\frac{K}{2}\sum_{n=1}^{N-1}\mathbf{Z}_n\mathbf{Z}_{n+1}\right)=\exp\left(\frac{K}{2}\sum_{n=1}^{N-1}\mathrm{i}\mathbf{\Gamma}_{2n}\mathbf{\Gamma}_{2n+1}\right), \tag{A3}$$

$$\mathbf{V}_2=\exp\left(K\sum_{n=1}^{N-1}\mathbf{Z}_n\mathbf{Z}_{n+1}\right)=\exp\left(K\sum_{n=1}^{N-1}\mathrm{i}\mathbf{\Gamma}_{2n}\mathbf{\Gamma}_{2n+1}\right),$$

$$\mathrm{e}^{-2K'}=\tanh\frac{J'}{kT},\qquad K=\frac{J}{kT}. \tag{A4}$$

In this paper by $\mathbf{I}_k$ we denote $k$ -order identity matrix. The forms of matrices $\mathbf{\Gamma}_n$ $(n=1,2,\cdots,2N)$ in (A3) read[4, 5]

$$\mathbf{\Gamma}_{2n-1}=\underbrace{\mathbf{X}\otimes\mathbf{X}\otimes\cdots\otimes\mathbf{X}}_{n-1}\otimes\mathbf{Z}\otimes\mathbf{I}_2\otimes\cdots\otimes\mathbf{I}_2\quad(n=1,2,\cdots,N),$$

$$\mathbf{\Gamma}_{2n}=\underbrace{\mathbf{X}\otimes\mathbf{X}\otimes\cdots\otimes\mathbf{X}}_{n-1}\otimes\mathbf{Y}\otimes\mathbf{I}_2\otimes\cdots\otimes\mathbf{I}_2\quad(n=1,2,\cdots,N); \tag{A5}$$

$$\mathbf{X}_n=\underbrace{\mathbf{I}_2\otimes\mathbf{I}_2\otimes\cdots\otimes\mathbf{I}_2}_{n-1}\otimes\mathbf{X}\otimes\mathbf{I}_2\otimes\cdots\otimes\mathbf{I}_2,$$

$$\mathbf{Y}_n=\underbrace{\mathbf{I}_2\otimes\mathbf{I}_2\otimes\cdots\otimes\mathbf{I}_2}_{n-1}\otimes\mathbf{Y}\otimes\mathbf{I}_2\otimes\cdots\otimes\mathbf{I}_2, \tag{A6}$$

$$\mathbf{Z}_n=\underbrace{\mathbf{I}_2\otimes\mathbf{I}_2\otimes\cdots\otimes\mathbf{I}_2}_{n-1}\otimes\mathbf{Z}\otimes\mathbf{I}_2\otimes\cdots\otimes\mathbf{I}_2\quad(n=1,2,\cdots,N);$$

$$\mathbf{X}=\begin{bmatrix}0&1\\1&0\end{bmatrix},\quad\mathbf{Y}=\begin{bmatrix}0&-\mathrm{i}\\\mathrm{i}&0\end{bmatrix},\quad\mathbf{Z}=\begin{bmatrix}1&0\\0&-1\end{bmatrix}. \tag{A7}$$

In this paper, by $[\mathbf{A},\mathbf{B}]$ and $\{\mathbf{A},\mathbf{B}\}$ we denote $\mathbf{AB}-\mathbf{BA}$ and $\mathbf{AB}+\mathbf{BA}$, respectively, for example, $\mathbf{\Gamma}_n$ $(n=1,2,\cdots,2N)$ given by (A5) satisfy

$$\{\mathbf{\Gamma}_l,\mathbf{\Gamma}_m\}=2\delta_{lm}\qquad(l,m=1,2,\cdots,2N). \tag{A8}$$

## A.2 Some results regarding the matrix $\mathbf{V}_2^{1/2}\mathbf{V}_1\mathbf{V}_2^{1/2}$

Both matrices $\mathbf{V}_1=R(\boldsymbol{\omega}_1)$ and $\mathbf{V}_2^{1/2}=R(\boldsymbol{\omega}_2)$ in (A2) are spin representatives[3], the concrete forms of the two orthogonal matrices $\boldsymbol{\omega}_1$ and $\boldsymbol{\omega}_2$ are written out in terms of [5]; and then, by means of the property of spin representative: $R(\boldsymbol{\mu})R(\mathbf{v})=R(\boldsymbol{\mu}\mathbf{v})$, we obtain

$$\mathbf{V}_2^{1/2}\mathbf{V}_1\mathbf{V}_2^{1/2}=R(\boldsymbol{\omega}_2)R(\boldsymbol{\omega}_1)R(\boldsymbol{\omega}_2)=R(\boldsymbol{\omega}_2\boldsymbol{\omega}_1\boldsymbol{\omega}_2), \tag{A9}$$

which is also a spin representatives.

In Ref.[4], an orthogonal matrix $\boldsymbol{\xi}$ has been found by generating function method such that

$$\boldsymbol{\xi}^{\mathrm{T}}(\boldsymbol{\omega}_2\boldsymbol{\omega}_1\boldsymbol{\omega}_2)\boldsymbol{\xi}=\boldsymbol{\lambda},\quad\boldsymbol{\lambda}=\mathrm{diag}\left[\boldsymbol{\lambda}_1\ \boldsymbol{\lambda}_2\ \cdots\ \boldsymbol{\lambda}_N\right],$$

$$\boldsymbol{\lambda}_n=\begin{bmatrix}\cosh\gamma_{n-1}&\mathrm{i}\sinh\gamma_{n-1}\\-\mathrm{i}\sinh\gamma_{n-1}&\cosh\gamma_{n-1}\end{bmatrix}\qquad(n=1,2,\cdots,N), \tag{A10}$$

where $\gamma_{n-1}$ $(n=1,2,\cdots,N)$ are determined by

$$\cosh\gamma_{n-1}=\cosh2K'\cosh2K-x_{n-1}\sinh2K'\sinh2K,\quad\gamma_{n-1}>0; \tag{A11}$$

where $x_{n-1}$ $(n=1,2,\cdots,N)$ are $N$ roots of the $N$-order algebraic equation in $x$ :

$$G_N(x)=0\,. \tag{A12}$$

The function $G_k(x)$ is defined by

$$G_k(x)=g_k(x)-2g_{k-1}(x)\coth 2K'\tanh K+g_{k-2}(x)\tanh^2 K \qquad (k=0,\pm1,\pm2,\pm3,\cdots)\,, \tag{A13}$$

where $g_n(x)$ is the Chebyshev polynomial of the second class:

$$g_n(x)=\sum_{k=0}^{[n/2]}\frac{(n+1)!}{(2k+1)!(n-2k)!}x^{n-2k}\left(x^2-1\right)^k\,, \tag{A14}$$

$g_n(x)$ is a $n$-degree polynomial in $x$ . Therefore, $G_N(x)$ in (A12) is a $N$-order polynomial in $x$ .

It has been proved in Refs.[4, 5] that the $N-1$ roots of the $N$-order algebraic equation in (A12) are

$$x_{n-1}=\cos\varphi_{n-1}\,,\quad \varphi_{n-1}=\frac{n-1}{N}\pi+\frac{\theta_{n-1}}{N}\qquad (n=2,3,\cdots,N)\,, \tag{A15}$$

where $\theta_{n-1}$ $(n=2,3,\cdots,N)$ are determined by

$$\begin{aligned}&\cosh 2K'\sinh 2K-\cos\varphi_{n-1}\sinh 2K'\cosh 2K=\sinh\gamma_{n-1}\cos\theta_{n-1}\,,\\ &\sin\varphi_{n-1}\sinh 2K'=\sinh\gamma_{n-1}\sin\theta_{n-1}\,,\qquad 0<\theta_{n-1}<\pi\,.\end{aligned} \tag{A16a}$$

or

$$\tan\theta_{n-1}=\frac{\sin\varphi_{n-1}\sinh 2K'}{\cosh 2K'\sinh 2K-\cos\varphi_{n-1}\sinh 2K'\cosh 2K}\,,\qquad 0<\theta_{n-1}<\pi\,. \tag{A16b}$$

If $N$ is finite, then solving for $\theta_{n-1}$ $(n=2,3,\cdots,N)$ will be complex; when $N$ is very large, the approximate values of $x_{n-1}$ and $\gamma_{n-1}$ $(n=2,3,\cdots,N)$ correcting to $\frac{1}{N}$ order given in Refs.[4, 5] are

$$x_{n-1}=\cos\frac{(n-1)\pi+\theta^{(0)}_{n-1}}{N}\,,\quad \gamma_{n-1}\approx\gamma^{(0)}_{n-1}+2\sin\frac{(n-1)\pi+\theta^{(0)}_{n-1}}{N}\sin\frac{\theta^{(0)}_{n-1}}{2N}\frac{\sinh 2K'\sinh 2K}{\sinh\gamma^{(0)}_{n-1}} \tag{A17}$$
$$(n=2,3,\cdots,N)\,,$$

where $\gamma^{(0)}_{n-1}$ and $\theta^{(0)}_{n-1}$ $(n=2,3,\cdots,N)$ are introduced by

$$\begin{aligned}&\cosh\gamma^{(0)}_{n-1}=\cosh 2K'\cosh 2K-\cos\frac{(n-1)\pi}{N}\sinh 2K'\sinh 2K\,,\qquad \gamma^{(0)}_{n-1}>0\,;\\ &\tan\theta^{(0)}_{n-1}=\frac{\sin\frac{(n-1)\pi}{N}\sinh 2K'}{\cosh 2K'\sinh 2K-\cos\frac{(n-1)\pi}{N}\sinh 2K'\cosh 2K}\,,\qquad 0<\theta^{(0)}_{n-1}<\pi\,.\end{aligned} \tag{A18}$$

The critical temperature $T_c$ is introduced in terms of

$$N=\frac{\sinh 2K'_c}{\sinh 2(K_c-K'_c)}\,, \tag{A19a}$$

$$e^{-2K'_c}=\tanh\frac{J'}{kT_c}\,,\qquad K_c=\frac{J}{kT_c}\,; \tag{A20}$$

substituting $K'_c$ and $K_c$ defined by (A20) into (A19a), (A19a) can be written in the form

$$\sinh\frac{2J}{kT_c}\cosh\frac{2J'}{kT_c}-\cosh\frac{2J}{kT_c}=\frac{1}{N}\,. \tag{A19b}$$

For $T\ge T_c$, the rest root $x_0$ of the $N$-th order algebraic equation in (A12) is still determined by (A15) and (A16b), and, satisfies $0 < x_0 \le 1$ ; but for $T < T_c$ , $K' < K$ and $x_0$ satisfies $1 < x_0 < \frac{\tanh 2K}{\tanh 2K'}$ . As $N \to \infty$ , the approximate values of $x_0$ and $\gamma_0$ given in Refs.[4, 5] are

$$x_0 \approx \begin{cases} \cos\frac{\pi}{N}, & T \ge T_c,\ K' > K; \\ \cos\frac{\pi}{2N}, & T \ge T_c,\ K' = K; \\ \frac{1}{2}\left(\frac{\tanh K}{\tanh K'} + \frac{\tanh K'}{\tanh K}\right) - 2\frac{(\cosh 2K - \cosh 2K')^2}{\sinh 2K' \sinh^3 2K}\left(\frac{\tanh K'}{\tanh K}\right)^{2N}, & T < T_{cc},\ K' < K, \end{cases} \tag{A21}$$

$$\gamma_0 \approx \begin{cases} 2(K'-K) + 2\sin^2\frac{\pi}{2N}\frac{\sinh 2K' \sinh 2K}{\sinh 2(K'-K)}, & T \ge T_c,\ K' > K; \\ 2\sin\frac{\pi}{4N}\sinh 2K, & T \ge T_c,\ K' = K; \\ 2\frac{\cosh 2K - \cosh 2K'}{\sinh 2K}\left(\frac{\tanh K'}{\tanh K}\right)^N \left(1 + \frac{\Pi}{\sinh^2 2K} N \left(\frac{\tanh K'}{\tanh K}\right)^{2N}\right), & T < T_c,\ K' < K, \end{cases} \tag{A22}$$

where

$$\Pi = 2(\cosh 2K - \cosh 2K') + \frac{1}{N}\left(\frac{1}{3}(\cosh 2K - \cosh 2K')^2 - \sinh^2 2K'\right).$$

As can be seen from (A22), for $T \ge T_c$ and $K' > K$ , $\gamma_0 \approx 2(K'-K)$ ; once the temperature $T$ of the system crosses the critical temperature $T_c$ , $K' < K$ and, thus, $\gamma_0$ becomes exponentially little and does vanish rapidly as $N \to \infty$ . This property of $\gamma_0$ plays a key role for properties of the correlation function $\langle \sigma_{1,1}\sigma_{1,N} \rangle$ given in Ref.[5], in this paper, which causes the $\Lambda_P$ defined by (54) to not be zero and, thus, leads to the appearance of the long range order (see (62)). Moreover, obtaining $\lim_{N\to\infty} e^{\pm\gamma_0} = 1$ from $\lim_{N\to\infty} \gamma_0 = 0$ means that the two eigenvalues $e^{\gamma_0}$ and $e^{-\gamma_0}$ of the matrix $\mathbf{V}_2^{1/2}\mathbf{V}_1\mathbf{V}_2^{1/2}$ defined by (A9) are equal, this degenerate eigenvalue leads to non-zero spontaneous magnetization.

The matrix $\xi$ in (A10) is expressed in terms of $2 \times 2$ blocks

$$\xi_{lm} = \begin{bmatrix} u_{lm} & 0 \\ 0 & v_{lm} \end{bmatrix}, \quad 1 \le l, m \le N, \tag{A23}$$

in which $u_{lm}$ and $v_{lm}$ are given by

$$\begin{aligned} u_{1m} &= \sqrt{2}\Omega_{m-1}, & v_{1m} &= \sqrt{2}\Omega_{m-1}\frac{\sinh\gamma_{m-1}}{\sinh 2K' \cosh K}; \\ u_{nm} &= \sqrt{2}\Omega_{m-1}G_{n-1}(x_{m-1})\cosh K, & v_{nm} &= \sqrt{2}\Omega_{m-1}\frac{\sinh\gamma_{m-1}g_{n-1}(x_{m-1})}{\sinh 2K' \cosh K}; \\ u_{Nm} &= \sqrt{2}\Omega_{m-1}G_{N-1}(x_{m-1})\cosh K, & v_{Nm} &= \sqrt{2}\Omega_{m-1}\frac{\sinh\gamma_{m-1}g_{N-1}(x_{m-1})}{\sinh 2K'} \end{aligned} \tag{A24}$$

$$(n = 2, 3, \cdots, N-1),$$

where the functions $G_k(x)$ and $g_k(x)$ are expressed by (A13) and (A14), respectively;

$$\Omega_{n-1} = \sinh 2K' \cosh K \sqrt{\frac{1 - x_{n-1}^2}{N \sinh^2 \gamma_{n-1} + \cosh \gamma_{n-1} \cosh 2K' - \cosh 2K}} . \tag{A25}$$

As $N \to \infty$, the approximate values of $\Omega_{n-1}$ given in Refs.[4, 5] are

$$\Omega_{n-1} \approx \frac{1}{\sqrt{N}} \sin \frac{(n-1)\pi}{N} \frac{\sinh 2K' \cosh K}{\sinh \gamma_{n-1}} \qquad (n = 2, 3, \cdots, N), \tag{A26}$$

$$\Omega_0^2 \approx \begin{cases} \dfrac{1}{N} \sin^2 \dfrac{\pi}{N} \dfrac{\sinh^2 2K' \cosh^2 K}{\sinh^2 2(K' - K)}, & T \geq T_{\mathrm{c}},\ K' > K; \\[2ex] \dfrac{1}{N} \dfrac{\sinh^2 2K' \cosh^2 K}{\sinh^2 2K}, & T \geq T_{\mathrm{c}},\ K' = K; \\[2ex] \dfrac{\cosh 2K - \cosh 2K'}{4 \sinh^2 K} + \Xi N \left( \dfrac{\tanh K'}{\tanh K} \right)^{2N}, & T < T_{\mathrm{c}},\ K' < K, \end{cases} \tag{A27}$$

where

$$\Xi = 4 \frac{\cosh 2K - \cosh 2K'}{\sinh^4 2K \sinh^2 2K'} \left( (\cosh 2K - \cosh 2K') - \frac{1}{N} \left( \frac{1}{2} \cosh 2K' (\cosh 2K + \cosh 2K') - 1 \right) \right) .$$

Since $\xi$ is orthogonal matrix, which satisfies $\xi^{\mathrm{T}} \xi = \mathbf{I}_{2N}$ and $\xi \xi^{\mathrm{T}} = \mathbf{I}_{2N}$. In terms of $\xi^{\mathrm{T}} \xi = \mathbf{I}_{2N}$ we obtain

$$\sum_{n=1}^{N} u_{nl} u_{nm} = \delta_{lm} , \qquad \sum_{n=1}^{N} v_{nl} v_{nm} = \delta_{lm} ; \tag{A28}$$

and, in terms of $\xi \xi^{\mathrm{T}} = \mathbf{I}_{2N}$ we obtain

$$\sum_{n=1}^{N} u_{ln} u_{mn} = \delta_{lm} , \qquad \sum_{n=1}^{N} v_{ln} v_{mn} = \delta_{lm} . \tag{A29}$$

**A.3 The calculation result of $Z_0$**

According to $R(\boldsymbol{\mu}) R(\mathbf{v}) = R(\boldsymbol{\mu}\mathbf{v})$, $\xi^{\mathrm{T}} \xi = \mathbf{I}_{2N}$ and $\xi \xi^{\mathrm{T}} = \mathbf{I}_{2N}$ we have

$$R(\xi^{\mathrm{T}}) R(\xi) = R(\xi^{\mathrm{T}} \xi) = R(\mathbf{I}_{2N}) = \mathbf{I}_{2^N} , \qquad R(\xi) R(\xi^{\mathrm{T}}) = R(\xi \xi^{\mathrm{T}}) = R(\mathbf{I}_{2N}) = \mathbf{I}_{2^N} . \tag{A30}$$

On the other hand, according to (A9), (A10) and $R(\boldsymbol{\mu}) R(\mathbf{v}) = R(\boldsymbol{\mu}\mathbf{v})$, we have

$$R(\xi^{\mathrm{T}}) \mathbf{V}_2^{1/2} \mathbf{V}_1 \mathbf{V}_2^{1/2} R(\xi) = R(\xi^{\mathrm{T}}) R(\boldsymbol{\omega}_2 \boldsymbol{\omega}_1 \boldsymbol{\omega}_2) R(\xi) = R(\xi^{\mathrm{T}} (\boldsymbol{\omega}_2 \boldsymbol{\omega}_1 \boldsymbol{\omega}_2) \xi) = R(\boldsymbol{\lambda}) ,$$

and, for the orthogonal matrix $\boldsymbol{\lambda}$ given by (A10), we have [5] $R(\boldsymbol{\lambda}) = \prod_{n=1}^{N} \mathrm{e}^{\frac{\gamma_{n-1}}{2} \boldsymbol{\Gamma}_{2n-1} \mathrm{i} \boldsymbol{\Gamma}_{2n}}$. Therefore,

$$\mathbf{V}_2^{1/2} \mathbf{V}_1 \mathbf{V}_2^{1/2} = R(\xi) \left( \prod_{n=1}^{N} \mathrm{e}^{\frac{\gamma_{n-1}}{2} \boldsymbol{\Gamma}_{2n-1} \mathrm{i} \boldsymbol{\Gamma}_{2n}} \right) R(\xi^{\mathrm{T}}) . \tag{A31}$$

And, further, using (A30) and (A31) we obtain

$$\left( \mathbf{V}_2^{1/2} \mathbf{V}_1 \mathbf{V}_2^{1/2} \right)^L = \left( R(\xi) \left( \prod_{n=1}^{N} \mathrm{e}^{\frac{\gamma_{n-1}}{2} \boldsymbol{\Gamma}_{2n-1} \mathrm{i} \boldsymbol{\Gamma}_{2n}} \right) R(\xi^{\mathrm{T}}) \right)^L = R(\xi) \left( \prod_{n=1}^{N} \mathrm{e}^{\frac{\gamma_{n-1}}{2} \boldsymbol{\Gamma}_{2n-1} \mathrm{i} \boldsymbol{\Gamma}_{2n}} \right)^L R(\xi^{\mathrm{T}}) . \tag{A32}$$

According to (A8), it is easy to prove

$$\left[\mathbf{\Gamma}_{2n-1}\mathbf{\Gamma}_{2n},\mathbf{\Gamma}_{2n'-1}\mathbf{\Gamma}_{2n'}\right]=0 \quad (n\neq n'), \tag{A33}$$

therefore, (A32) can be written in the form

$$\left(\mathbf{V}_2^{1/2}\mathbf{V}_1\mathbf{V}_2^{1/2}\right)^L = R(\xi)\left(\prod_{n=1}^{N} \mathrm{e}^{\frac{L\gamma_{n-1}}{2}\mathbf{\Gamma}_{2n-1}\mathrm{i}\mathbf{\Gamma}_{2n}}\right)R\left(\xi^{\mathrm{T}}\right). \tag{A34}$$

From the above derivation process we see that (A34) holds for arbitrary positive integer $L$.

Substituting (A34) into (A2), using (A30) and the property of trace: $\mathrm{Tr}(\mathbf{AB})=\mathrm{Tr}(\mathbf{BA})$, we obtain

$$\begin{aligned}Z_0 &= \mathrm{Tr}\left(\left(\mathbf{V}_2^{1/2}\mathbf{V}_1\mathbf{V}_2^{1/2}\right)^L\right)=\mathrm{Tr}\left(R(\xi)\left(\prod_{n=1}^{N} \mathrm{e}^{\frac{L\gamma_{n-1}}{2}\mathbf{\Gamma}_{2n-1}\mathrm{i}\mathbf{\Gamma}_{2n}}\right)R\left(\xi^{\mathrm{T}}\right)\right)=\mathrm{Tr}\left(R\left(\xi^{\mathrm{T}}\right)R(\xi)\prod_{n=1}^{N} \mathrm{e}^{\frac{L\gamma_{n-1}}{2}\mathbf{\Gamma}_{2n-1}\mathrm{i}\mathbf{\Gamma}_{2n}}\right)\\ &=\mathrm{Tr}\left(\mathbf{I}_{2^N}\prod_{n=1}^{N} \mathrm{e}^{\frac{L\gamma_{n-1}}{2}\mathbf{\Gamma}_{2n-1}\mathrm{i}\mathbf{\Gamma}_{2n}}\right)=\mathrm{Tr}\left(\prod_{n=1}^{N} \mathrm{e}^{\frac{L\gamma_{n-1}}{2}\mathbf{\Gamma}_{2n-1}\mathrm{i}\mathbf{\Gamma}_{2n}}\right).\end{aligned} \tag{A35}$$

Defining $\boldsymbol{\eta}=\frac{1}{\sqrt{2}}\begin{bmatrix}1 & -1\\ 1 & 1\end{bmatrix}$, we have $\boldsymbol{\eta}^{-1}=\frac{1}{\sqrt{2}}\begin{bmatrix}1 & 1\\ -1 & 1\end{bmatrix}$ and $\boldsymbol{\eta}^{-1}\mathbf{X}\boldsymbol{\eta}=\mathbf{Z}$, where both $\mathbf{X}$ and $\mathbf{Z}$ are given by (A7); and, further, defining $\boldsymbol{\Delta}=\underbrace{\boldsymbol{\eta}\otimes\boldsymbol{\eta}\otimes\cdots\otimes\boldsymbol{\eta}}_{N}$, we have $\boldsymbol{\Delta}^{-1}=\underbrace{\boldsymbol{\eta}^{-1}\otimes\boldsymbol{\eta}^{-1}\otimes\cdots\otimes\boldsymbol{\eta}^{-1}}_{N}$ and $\boldsymbol{\Delta}^{-1}\mathbf{\Gamma}_{2n-1}\mathrm{i}\mathbf{\Gamma}_{2n}\boldsymbol{\Delta}=\boldsymbol{\Delta}^{-1}\mathbf{X}_n\boldsymbol{\Delta}=\mathbf{Z}_n$. As can be seen from (A6), all $\mathbf{Z}_n$ are diagonal matrices, hence, in terms of $\mathrm{Tr}(\mathbf{AB})=\mathrm{Tr}(\mathbf{BA})$, from (A35) we obtain

$$\begin{aligned}Z_0 &= \mathrm{Tr}\left(\prod_{n=1}^{N} \mathrm{e}^{\frac{L\gamma_{n-1}}{2}\mathbf{\Gamma}_{2n-1}\mathrm{i}\mathbf{\Gamma}_{2n}}\right)=\mathrm{Tr}\left(\mathbf{I}_{2^N}\prod_{n=1}^{N} \mathrm{e}^{\frac{L\gamma_{n-1}}{2}\mathbf{\Gamma}_{2n-1}\mathrm{i}\mathbf{\Gamma}_{2n}}\right)=\mathrm{Tr}\left(\boldsymbol{\Delta}\boldsymbol{\Delta}^{-1}\prod_{n=1}^{N} \mathrm{e}^{\frac{L\gamma_{n-1}}{2}\mathbf{\Gamma}_{2n-1}\mathrm{i}\mathbf{\Gamma}_{2n}}\right)\\ &=\mathrm{Tr}\left(\boldsymbol{\Delta}^{-1}\left(\prod_{n=1}^{N} \mathrm{e}^{\frac{L\gamma_{n-1}}{2}\mathbf{\Gamma}_{2n-1}\mathrm{i}\mathbf{\Gamma}_{2n}}\right)\boldsymbol{\Delta}\right)=\mathrm{Tr}\left(\prod_{n=1}^{N}\boldsymbol{\Delta}^{-1} \mathrm{e}^{\frac{L\gamma_{n-1}}{2}\mathbf{\Gamma}_{2n-1}\mathrm{i}\mathbf{\Gamma}_{2n}}\boldsymbol{\Delta}\right)=\mathrm{Tr}\left(\prod_{n=1}^{N} \mathrm{e}^{\frac{L\gamma_{n-1}}{2}\boldsymbol{\Delta}^{-1}\mathbf{\Gamma}_{2n-1}\mathrm{i}\mathbf{\Gamma}_{2n}\boldsymbol{\Delta}}\right)\\ &=\mathrm{Tr}\left(\prod_{n=1}^{N} \mathrm{e}^{\frac{L\gamma_{n-1}}{2}\mathbf{Z}_n}\right)=\sum \mathrm{e}^{\pm\frac{L\gamma_0}{2}}\mathrm{e}^{\pm\frac{L\gamma_1}{2}}\cdots\mathrm{e}^{\pm\frac{L\gamma_{N-1}}{2}}=\prod_{n=1}^{N}\left(2\cosh\frac{L\gamma_{n-1}}{2}\right).\end{aligned} \tag{A36}$$

Eq.(A36) is a well-known result, what we want to emphasize is that, according to the above derivation process, (A36) holds for arbitrary positive integer $L$ and arbitrary values of $\gamma_{n-1}$ $(n=1,2,\cdots,N)$.

## Appendix B Expressions of $\langle\sigma_{1,1}\sigma_{1+P,1}\rangle$ in the thermodynamic limit given by Ref.[1]

In this Appendix, we prove that expressions of $\langle\sigma_{1,1}\sigma_{1+P,1}\rangle$ in the thermodynamic limit and under the limit order $\lim_{L\to\infty}\lim_{N\to\infty}\langle\sigma_{1,1}\sigma_{1+P,1}\rangle$ given as (49) and (67) are the same as that given by Ref.[1].

In Ref.[1], the partition function of the two-dimensional rectangular Ising model with periodic-free boundary conditions in presence of boundary magnetic field $B$ is given by (VI.2.2), i.e., the formula (2.2) of Chapter VI of Ref.[1] (the same below):

$$Z=\sum_{\sigma=\pm1}\exp\left(\beta E_1\sum_{j=1}^{2M}\sum_{k=-(N-1)}^{N}\sigma_{j,k}\sigma_{j,k+1}+\beta E_2\sum_{j=1}^{2M-1}\sum_{k=-(N-1)}^{N}\sigma_{j,k}\sigma_{j+1,k}+\beta B\sum_{k=-(N-1)}^{N}\sigma_{1,k}\right), \tag{B1}$$

with $k=N+1$ identified with $k=-(N-1)$.

Comparing (B1) with (2), since $\beta=\dfrac{1}{kT}$ (see (II.3.41)), we see that $E_1$ and $E_2$ in (B1) are denoted as $J'$ and $J$ in this paper, respectively, in this paper we only consider the case of $J'>0$ and $J>0$; some quantities in Ref.[1] can thus be denoted by using symbols of this paper. For example, $z_1$ and $z_2$ given by (V.2.5) can be written in the form:

$$z_1=\tanh\frac{J'}{kT}=\mathrm{e}^{-2K'},\quad z_2=\tanh\frac{J}{kT}=\tanh K; \tag{B2}$$

the boundary spontaneous magnetization $M_1(0^+)$ given by (VI.5.19) and (VI.5.20) can be written in the form:

$$M_1(0^+)=\frac{1}{2z_2}\sqrt{\frac{z_2^2(1+z_1)^2-(1-z_1)^2}{z_1}}=\sqrt{\frac{\cosh\frac{2J}{kT}-\coth\frac{2J'}{kT}}{\cosh\frac{2J}{kT}-1}}=\frac{\sqrt{\cosh 2K-\cosh 2K'}}{\sqrt{2}\sinh K}. \tag{B3}$$

**B.1 Correlation functions between two spins in the boundary row given by Ref.[1]**

Comparing (B1) with (2), we see that $\sigma_{l,m}$ in Ref.[1] is denoted as $\sigma_{m,l}$ in this paper; the definition of correlation functions $S_{11}(n,B)=\langle\sigma_{1,0}\sigma_{1,n}\rangle$ between two spins in the boundary row is given by (VII.2.1); as boundary magnetic field $B\to 0$, $S_{11}(n,0)$ in the thermodynamic limit is given as (VII.4.16a) and (VII.4.16b) for $T>T_\mathrm{c}$ and $T<T_\mathrm{c}$, respectively, i.e.,

$$S_{11}(n,0)=-\frac{1-z_1^2}{z_1z_2}\frac{1}{2\pi\mathrm{i}}\int_\Gamma \mathrm{d}\zeta\frac{\zeta^n}{\left(\zeta^2-1\right)\alpha}\qquad \text{for } T>T_\mathrm{c}, \tag{B4}$$

$$S_{11}(n,0)=M_1^2(0^+)-\frac{1-z_1^2}{z_1z_2}\frac{1}{2\pi\mathrm{i}}\int_\Gamma \mathrm{d}\zeta\frac{\zeta^n}{\left(\zeta^2-1\right)\alpha}\qquad \text{for } T<T_\mathrm{c}. \tag{B5}$$

In the above two formulas, $\zeta=\mathrm{e}^{\mathrm{i}\theta}$, both $z_1$ and $z_2$ are given as (B2), the contour $\Gamma$ goes around the branch-cuts of $\alpha$ inside the unit circle; according to the expression of $\alpha$ given as (VI.3.16), $\alpha^{\pm1}$ can be written in the form:

$$\alpha^{\pm1}=\beta_1\pm\beta_2; \tag{B6}$$

$$\beta_1=\frac{\left(1+z_1^2\right)\left(1+z_2^2\right)-z_1\left(1-z_2^2\right)\left(\mathrm{e}^{\mathrm{i}\theta}+\mathrm{e}^{-\mathrm{i}\theta}\right)}{2z_2\left(1-z_1^2\right)}=\frac{\left(1+z_1^2\right)\left(1+z_2^2\right)-z_1\left(1-z_2^2\right)\left(\zeta+\zeta^{-1}\right)}{2z_2\left(1-z_1^2\right)}, \tag{B7}$$

$$\begin{aligned}\beta_2=\sqrt{\beta_1^2-1}&=\frac{1-z_2^2}{2z_2\left(1-z_1^2\right)}\sqrt{\left(1-\alpha_1\mathrm{e}^{\mathrm{i}\theta}\right)\left(1-\alpha_1\mathrm{e}^{-\mathrm{i}\theta}\right)\left(1-\alpha_2^{-1}\mathrm{e}^{\mathrm{i}\theta}\right)\left(1-\alpha_2^{-1}\mathrm{e}^{-\mathrm{i}\theta}\right)}\\&=\frac{1-z_2^2}{2z_2\left(1-z_1^2\right)}\sqrt{\frac{\alpha_1}{\alpha_2}}\frac{\sqrt{(\zeta-\alpha_1)\left(\zeta-\alpha_1^{-1}\right)(\zeta-\alpha_2)\left(\zeta-\alpha_2^{-1}\right)}}{\zeta},\end{aligned} \tag{B8}$$

where both $\alpha_1$ and $\alpha_2$ are given as (VI.3.17), i.e.,

$$\alpha_1 = \frac{z_1(1-z_2)}{1+z_2} = \mathrm{e}^{-2(K'+K)}, \qquad \alpha_2 = \frac{1-z_2}{z_1(1+z_2)} = \mathrm{e}^{2(K'-K)}, \tag{B9}$$

where we have used (B2), from (B9) we have $\alpha_1 > 0$, $\alpha_2 > 0$ and $\dfrac{\alpha_1}{\alpha_2} = \mathrm{e}^{-4K'} > 0$.

According to (B2), for $\dfrac{1-z_2^2}{z_2\left(1-z_1^2\right)}$ in (B8), we have $\dfrac{1-z_2^2}{z_2\left(1-z_1^2\right)} = \dfrac{\mathrm{e}^{2K'}}{\sinh 2K' \sinh 2K} > 0$.

In Ref.[1], a different expression of $S_{11}(n,0)$ for $T > T_\mathrm{c}$ is presented by (VII.8.39), i.e.,

$$S_{11}(n,0) = -\frac{1-z_2^2}{z_1 z_2^2}\frac{1}{2\pi}\int_{\alpha_1}^{\alpha_2^{-1}} \mathrm{d}\zeta \frac{\zeta^n}{\zeta^2-1}\sqrt{\left(1-\alpha_1\zeta\right)\left(1-\alpha_1\zeta^{-1}\right)\left(1-\alpha_2^{-1}\zeta\right)\left(\alpha_2^{-1}\zeta^{-1}-1\right)} \quad \text{for } T > T_\mathrm{c}; \tag{B10}$$

however, no derivation process from (B4) to (B10) is given in Ref.[1], here we present the derivation process.

According to (B6) ~ (B8), in the $\zeta$ -plane, $\alpha$ has four branch points, at $\alpha_1^{\pm 1}$ and $\alpha_2^{\pm 1}$. Since $K' > K$ holds true when $T > T_\mathrm{c}$ holds true, and $K' < K$ holds true when $T < T_\mathrm{c}$ holds true, in terms of (B9) we have

$$0 < \alpha_1 = \mathrm{e}^{-2(K'+K)} < \alpha_2^{-1} = \mathrm{e}^{-2(K'-K)} < 1 < \alpha_2 = \mathrm{e}^{2(K'-K)} < \alpha_1^{-1} = \mathrm{e}^{-2(K'+K)} \quad \text{for } T > T_\mathrm{c},$$

$$0 < \alpha_1 = \mathrm{e}^{-2(K'+K)} < \alpha_2 = \mathrm{e}^{-2(K-K')} < 1 < \alpha_2^{-1} = \mathrm{e}^{2(K-K')} < \alpha_1^{-1} = \mathrm{e}^{-2(K'+K)} \quad \text{for } T < T_\mathrm{c}.$$

Since the contour $\Gamma$ in either (B4) or (B5) goes around the branch-cuts of $\alpha$ inside the unit circle, which thus be illustrated in Fig.B1 and Fig.B2, respectively.

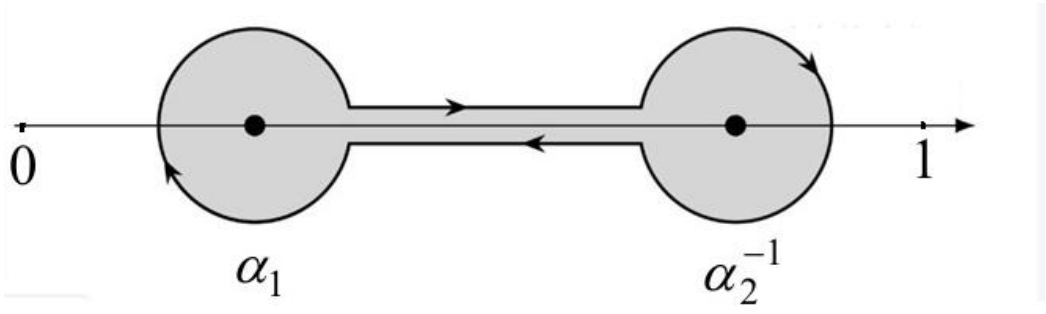

Fig.B1 The contour $\Gamma$ for $T > T_\mathrm{c}$

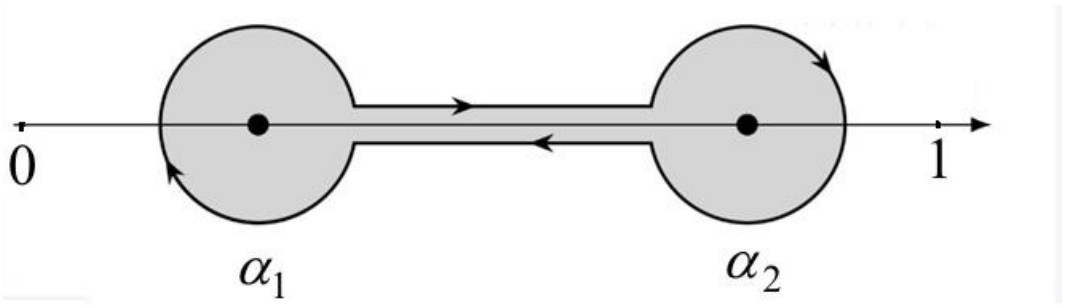

Fig.B2 The contour $\Gamma$ for $T < T_\mathrm{c}$

Substituting $\alpha^{-1} = \beta_1 - \beta_2$ given as (B6) into (B4), we have

$$S_{1,1}(n,0) = -\frac{1-z_1^2}{z_1 z_2}\frac{1}{2\pi\mathrm{i}}\int_\Gamma \mathrm{d}\zeta \frac{\zeta^n}{\zeta^2-1}\beta_1 + \frac{1-z_1^2}{z_1 z_2}\frac{1}{2\pi\mathrm{i}}\int_\Gamma \mathrm{d}\zeta \frac{\zeta^n}{\zeta^2-1}\beta_2; \tag{B11}$$

According to (B7) and Fig.B1, the integrand $\dfrac{\zeta^n}{\zeta^2-1}\beta_1$ in (B11) is normal in the area surrounded by the contour $\Gamma$, therefore, $\displaystyle\int_\Gamma \mathrm{d}\zeta \frac{\zeta^n}{\zeta^2-1}\beta_1 = 0$, and, further, using the expression of $\beta_2$ given as (B8), (B11) becomes

$$
\begin{aligned}
S_{1,1}(n,0) &= \frac{1-z_2^2}{2z_1z_2^2}\sqrt{\frac{\alpha_1}{\alpha_2}}\frac{1}{2\pi\mathrm{i}}\int_\Gamma \mathrm{d}\zeta \frac{\zeta^n}{\zeta^2-1}\frac{\sqrt{(\zeta-\alpha_1)\left(\zeta-\alpha_1^{-1}\right)(\zeta-\alpha_2)\left(\zeta-\alpha_2^{-1}\right)}}{\zeta} \\
&= \frac{1-z_2^2}{2z_1z_2^2}\sqrt{\frac{\alpha_1}{\alpha_2}}\frac{1}{2\pi\mathrm{i}}\int_{\alpha_1}^{\alpha_2^{-1}} \mathrm{d}\zeta \frac{\zeta^n}{\zeta^2-1}\frac{\sqrt{|\zeta-\alpha_1|\mathrm{e}^{\mathrm{i}\times 0\pi}\times\left|\zeta-\alpha_1^{-1}\right|\mathrm{e}^{\mathrm{i}\pi}\times|\zeta-\alpha_2|\mathrm{e}^{\mathrm{i}\pi}\times\left|\zeta-\alpha_2^{-1}\right|\mathrm{e}^{\mathrm{i}\pi}}}{\zeta} \\
&+ \frac{1-z_2^2}{2z_1z_2^2}\sqrt{\frac{\alpha_1}{\alpha_2}}\frac{1}{2\pi\mathrm{i}}\int_{\alpha_2^{-1}}^{\alpha_1} \mathrm{d}\zeta \frac{\zeta^n}{\zeta^2-1}\frac{\sqrt{|\zeta-\alpha_1|\mathrm{e}^{\mathrm{i}\times 2\pi}\times\left|\zeta-\alpha_1^{-1}\right|\mathrm{e}^{\mathrm{i}\pi}\times|\zeta-\alpha_2|\mathrm{e}^{\mathrm{i}\pi}\times\left|\zeta-\alpha_2^{-1}\right|\mathrm{e}^{\mathrm{i}\pi}}}{\zeta} \\
&= \frac{1-z_2^2}{2z_1z_2^2}\sqrt{\frac{\alpha_1}{\alpha_2}}\frac{1}{2\pi\mathrm{i}}\int_{\alpha_1}^{\alpha_2^{-1}} \mathrm{d}\zeta \frac{\zeta^n}{\zeta^2-1}\mathrm{e}^{\mathrm{i}\frac{3\pi}{2}}\frac{\sqrt{|\zeta-\alpha_1|\left|\zeta-\alpha_1^{-1}\right||\zeta-\alpha_2|\left|\zeta-\alpha_2^{-1}\right|}}{\zeta} \\
&+ \frac{1-z_2^2}{2z_1z_2^2}\sqrt{\frac{\alpha_1}{\alpha_2}}\frac{1}{2\pi\mathrm{i}}\int_{\alpha_2^{-1}}^{\alpha_1} \mathrm{d}\zeta \frac{\zeta^n}{\zeta^2-1}\mathrm{e}^{\mathrm{i}\frac{5\pi}{2}}\frac{\sqrt{|\zeta-\alpha_1|\left|\zeta-\alpha_1^{-1}\right||\zeta-\alpha_2|\left|\zeta-\alpha_2^{-1}\right|}}{\zeta} \\
&= -\frac{1-z_2^2}{z_1z_2^2}\sqrt{\frac{\alpha_1}{\alpha_2}}\frac{1}{2\pi}\int_{\alpha_1}^{\alpha_2^{-1}} \mathrm{d}\zeta \frac{\zeta^n}{\zeta^2-1}\frac{\sqrt{|\zeta-\alpha_1|\left|\zeta-\alpha_1^{-1}\right||\zeta-\alpha_2|\left|\zeta-\alpha_2^{-1}\right|}}{\zeta}.
\end{aligned} \tag{B12}
$$

On the other hand, using (B9), by straightforward calculation we obtain

$$
\begin{aligned}
&\frac{1-z_2^2}{2z_2\left(1-z_1^2\right)}\sqrt{\frac{\alpha_1}{\alpha_2}}\frac{\sqrt{|\zeta-\alpha_1|\left|\zeta-\alpha_1^{-1}\right||\zeta-\alpha_2|\left|\zeta-\alpha_2^{-1}\right|}}{\zeta} \\
&= \sqrt{\left|1-\left[\frac{\left(1+z_1^2\right)\left(1+z_2^2\right)-z_1\left(1-z_2^2\right)\left(\zeta+\zeta^{-1}\right)}{2z_2\left(1-z_1^2\right)}\right]^2\right|} \\
&= \sqrt{|f_1(\zeta)||f_2(\zeta)|};
\end{aligned} \tag{B13}
$$

$$
\begin{aligned}
f_1(\zeta) &= 1-\frac{\left(1+z_1^2\right)\left(1+z_2^2\right)-z_1\left(1-z_2^2\right)\left(\zeta+\zeta^{-1}\right)}{2z_2\left(1-z_1^2\right)} \\
&= \frac{2z_2\left(1-z_1^2\right)-\left(1+z_1^2\right)\left(1+z_2^2\right)+z_1\left(1-z_2^2\right)f(\zeta)}{2z_2\left(1-z_1^2\right)},
\end{aligned} \tag{B14}
$$

$$
\begin{aligned}
f_2(\zeta) &= 1+\frac{\left(1+z_1^2\right)\left(1+z_2^2\right)-z_1\left(1-z_2^2\right)\left(\zeta+\zeta^{-1}\right)}{2z_2\left(1-z_1^2\right)} \\
&= \frac{2z_2\left(1-z_1^2\right)+\left(1+z_1^2\right)\left(1+z_2^2\right)-z_1\left(1-z_2^2\right)f(\zeta)}{2z_2\left(1-z_1^2\right)},
\end{aligned} \tag{B15}
$$

in (B14) and (B15),

$$
f(\zeta) = \zeta + \zeta^{-1}. \tag{B16}
$$

For the function $f(\zeta)$ given as (B16), we have $\frac{\mathrm{d}f(\zeta)}{\mathrm{d}\zeta} = -\frac{1-\zeta^2}{\zeta^2}$, therefore, in the area $0 < \alpha_1 \le \zeta \le \alpha_2^{-1} < 1$, $f(\zeta)$ is a strictly monotone decreasing function as $\zeta$ increases, whose minimum and maximum in the area are thus

$$f_{\min}=f\left(\alpha_2^{-1}\right)=\alpha_2^{-1}+a_2=2\cosh 2(K'-K)\,,\qquad f_{\max}=f\left(\alpha_1\right)=\alpha_1+\alpha_1^{-1}=2\cosh 2(K'+K)\,. \tag{B17}$$

Therefore, for the functions $f_1(\zeta)$ and $f_2(\zeta)$ given as (B14) and (B15) respectively, we obtain

$$\begin{aligned} f_1(\zeta) &\geq \frac{2z_2\left(1-z_1^2\right)-\left(1+z_1^2\right)\left(1+z_2^2\right)+z_1\left(1-z_2^2\right)f_{\min}}{2z_2\left(1-z_1^2\right)} \\ &=\frac{2z_2\left(1-z_1^2\right)-\left(1+z_1^2\right)\left(1+z_2^2\right)+z_1\left(1-z_2^2\right)\left(\alpha_2^{-1}+a_2\right)}{2z_2\left(1-z_1^2\right)}=0\,, \end{aligned} \tag{B18}$$

$$\begin{aligned} f_2(\zeta) &\geq 1+\frac{\left(1+z_1^2\right)\left(1+z_2^2\right)-z_1\left(1-z_2^2\right)f_{\max}}{2z_2\left(1-z_1^2\right)} \\ &=\frac{2z_2\left(1-z_1^2\right)+\left(1+z_1^2\right)\left(1+z_2^2\right)-z_1\left(1-z_2^2\right)\left(\alpha_1+\alpha_1^{-1}\right)}{2z_2\left(1-z_1^2\right)}=0\,, \end{aligned} \tag{B19}$$

in the calculation process of (B18) and (B19), we have used (B9).

According to (B18) and (B19), both $f_1(\zeta)$ and $f_2(\zeta)$ are greater than zero in the area $0<\alpha_1\leq\zeta\leq\alpha_2^{-1}<1$, (B13) can thus be written in the form

$$\begin{aligned} &\frac{1-z_2^2}{2z_2\left(1-z_1^2\right)}\sqrt{\frac{\alpha_1}{\alpha_2}}\,\frac{\sqrt{\left|\zeta-\alpha_1\right|\left|\zeta-\alpha_1^{-1}\right|\left|\zeta-\alpha_2\right|\left|\zeta-\alpha_2^{-1}\right|}}{\zeta} \\ &=\sqrt{f_1(\zeta)f_2(\zeta)}=\sqrt{1-\left[\frac{\left(1+z_1^2\right)\left(1+z_2^2\right)-z_1\left(1-z_2^2\right)\left(\zeta+\zeta^{-1}\right)}{2z_2\left(1-z_1^2\right)}\right]^2}\,. \end{aligned} \tag{B20}$$

Substituting (B20) into (B12), we obtain

$$S_{1,1}(n,0)=-\frac{1-z_1^2}{z_1z_2}\frac{1}{\pi}\int_{\alpha_1}^{\alpha_2^{-1}}\mathrm{d}\zeta\,\frac{\zeta^n}{\zeta^2-1}\sqrt{1-\left[\frac{\left(1+z_1^2\right)\left(1+z_2^2\right)-z_1\left(1-z_2^2\right)\left(\zeta+\zeta^{-1}\right)}{2z_2\left(1-z_1^2\right)}\right]^2}\,. \tag{B21}$$

On the other hand, for the term $\sqrt{\left(1-\alpha_1\zeta\right)\left(1-\alpha_1\zeta^{-1}\right)\left(1-\alpha_2^{-1}\zeta\right)\left(\alpha_2^{-1}\zeta^{-1}-1\right)}$ in (B10), using (B9), by straightforward calculation we obtain

$$\begin{aligned} &\frac{1-z_2^2}{2z_2\left(1-z_1^2\right)}\sqrt{\left(1-\alpha_1\zeta\right)\left(1-\alpha_1\zeta^{-1}\right)\left(1-\alpha_2^{-1}\zeta\right)\left(\alpha_2^{-1}\zeta^{-1}-1\right)} \\ &=\sqrt{1-\left[\frac{\left(1+z_1^2\right)\left(1+z_2^2\right)-z_1\left(1-z_2^2\right)\left(\zeta+\zeta^{-1}\right)}{2z_2\left(1-z_1^2\right)}\right]^2}\,; \end{aligned}$$

substituting the above result into (B10), what we obtain is just (B21). We therefore have obtained (B10) from (B4).

For $T<T_{\mathrm{c}}$, the contour $\Gamma$ in (B5) is illustrated in Fig.B2; similar to the above calculation process, $S_{1,1}(n,0)$ given by (B5) can be written in the form

$$S_{1,1}(n,0)=M_1^2(0^+)-\frac{1-z_1^2}{z_1z_2}\frac{1}{\pi}\int_{\alpha_1}^{\alpha_2}\mathrm{d}\zeta\,\frac{\zeta^n}{\zeta^2-1}\sqrt{1-\left[\frac{\left(1+z_1^2\right)\left(1+z_2^2\right)-z_1\left(1-z_2^2\right)\left(\zeta+\zeta^{-1}\right)}{2z_2\left(1-z_1^2\right)}\right]^2}\,. \tag{B22}$$

**B.2 Variable substitution in the two integrals in (B21) and (B22)**

The variable $\zeta$ in the integral in (B21) satisfies $0 < \alpha_1 \le \zeta \le \alpha_2^{-1} < 1$, if we make variable substitution

$$\zeta = \zeta(\omega) = \mathrm{e}^{-\gamma(\omega)} = \cosh\gamma(\omega) - \sinh\gamma(\omega) = \cosh\gamma(\omega) - \sqrt{\cosh^2\gamma(\omega) - 1}\,, \tag{B23}$$

where $\cosh\gamma(\omega)$ is given as (49), then in terms of (49) and (B23), we have

$$\zeta = \begin{cases} \alpha_1 = \mathrm{e}^{-2(K'+K)}, & \omega = \pi\,; \\ \alpha_2^{-1} = \mathrm{e}^{-2(K'-K)}, & \omega = 0\,. \end{cases} \tag{B24}$$

The function $f(\zeta)$ given as (B16) can be written in the form

$$f(\zeta(\omega)) = \zeta(\omega) + (\zeta(\omega))^{-1} = \mathrm{e}^{-\gamma(\omega)} + \mathrm{e}^{\gamma(\omega)} = 2\cosh\gamma(\omega)\,, \tag{B25}$$

which is a strictly monotone increasing function as $\omega$ increases in the area $0 \le \omega \le \pi$, and reaches minimum and maximum at $\omega = 0$ and $\omega = \pi$, respectively; and, further, in terms of (49), it is easy to prove that the minimum and maximum at $\omega = 0$ and $\omega = \pi$ respectively of the function $f(\zeta(\omega))$ are just that given as (B17).

The above properties explain why the variable substitution given as (B23) is reasonable for the integral in (B21). Hence, after making such variable substitution, using (B23), (B24) and (B25), (B21) becomes

$$S_{1,1}(n,0) = -\frac{1-z_1^2}{z_1 z_2}\frac{1}{\pi}\int_{\pi}^{0}\frac{\mathrm{d}\zeta(\omega)}{\mathrm{d}\omega}\mathrm{d}\omega\frac{\mathrm{e}^{-n\gamma(\omega)}}{(\zeta(\omega))^2-1}\sqrt{1-\left[\frac{\left(1+z_1^2\right)\left(1+z_2^2\right)-2z_1\left(1-z_2^2\right)\cosh\gamma(\omega)}{2z_2\left(1-z_1^2\right)}\right]^2}\,. \tag{B26}$$

The variable $\zeta$ in the integral in (B22) satisfies $0 < \alpha_1 \le \zeta \le \alpha_2 < 1$, if we make the variable substitution given as (B23), then in terms of (49) and (B23), we have

$$\zeta = \begin{cases} \alpha_1 = \mathrm{e}^{-2(K+K')}, & \omega = \pi\,; \\ \alpha_2 = \mathrm{e}^{-2(K-K')}, & \omega = 0\,. \end{cases} \tag{B27}$$

And, similarly, by investigating some properties of the function $f(\zeta)$ given as (B16), we conclude that the variable substitution given as (B23) is reasonable for the integral in (B22). Hence, after making such variable substitution, using (B23), (B25) and (B27), (B22) becomes

$$S_{1,1}(n,0) = M_1^2(0^+) - \frac{1-z_1^2}{z_1 z_2}\frac{1}{\pi}\int_{\pi}^{0}\frac{\mathrm{d}\zeta(\omega)}{\mathrm{d}\omega}\mathrm{d}\omega\frac{\mathrm{e}^{-n\gamma(\omega)}}{(\zeta(\omega))^2-1}\sqrt{1-\left[\frac{\left(1+z_1^2\right)\left(1+z_2^2\right)-2z_1\left(1-z_2^2\right)\cosh\gamma(\omega)}{2z_2\left(1-z_1^2\right)}\right]^2}\,. \tag{B28}$$

As can be seen from (B26) and (B28), both integrals in (B26) and (B28) are the same.

In terms of (49), (B2) and (B23), by straightforward calculation we obtain

$$\frac{1-z_1^2}{z_1 z_2} = \frac{2\sinh 2K'}{\tanh K}\,, \qquad \frac{1}{(\zeta(\omega))^2-1} = \frac{1}{\left(\mathrm{e}^{-\gamma(\omega)}\right)^2-1} = -\frac{\mathrm{e}^{\gamma(\omega)}}{2\sinh\gamma(\omega)}\,,$$

$$\frac{\mathrm{d}\zeta(\omega)}{\mathrm{d}\omega} = \frac{\mathrm{d}\cosh\gamma(\omega)}{\mathrm{d}\omega} - \frac{\mathrm{d}\sqrt{\cosh^2\gamma(\omega)-1}}{\mathrm{d}\omega} = -\sinh 2K'\sinh 2K\frac{\sin\omega}{\sinh\gamma(\omega)}\mathrm{e}^{-\gamma(\omega)}\,,$$

$$\sqrt{1-\left[\frac{\left(1+z_1^2\right)\left(1+z_2^2\right)-2z_1\left(1-z_2^2\right)\cosh\gamma(\omega)}{2z_2\left(1-z_1^2\right)}\right]^2}=\sqrt{1-\left(\frac{\cosh 2K'\cosh 2K-\cosh\gamma(\omega)}{\sinh 2K'\sinh 2K}\right)^2}$$

$$=\sqrt{1-\left(\frac{\cosh 2K'\cosh 2K-(\cosh 2K'\cosh 2K-\sinh 2K'\sinh 2K\cos\omega)}{\sinh 2K'\sinh 2K}\right)^2}$$

$$=\sqrt{1-\cos^2\omega}=\sqrt{\sin^2\omega}=\sin\omega\,,$$

in the last step of the calculation process in the above formula, we have used the property $\sin\omega\ge 0$ in the area $0\le\omega\le\pi$.

Substituting the above four results into (B26) and (B28), we obtain

$$S_{1,1}(n,0)=2\sinh^2 2K'\cosh^2 K\int_0^{\pi}\frac{\mathrm{d}\omega}{\pi}\frac{\sin^2\omega}{\sinh^2\gamma(\omega)}\mathrm{e}^{-n\gamma(\omega)}=\Phi(n)\,, \tag{B29}$$

$$S_{1,1}(n,0)=M_1^2(0^+)+2\sinh^2 2K'\cosh^2 K\int_0^{\pi}\frac{\mathrm{d}\omega}{\pi}\frac{\sin^2\omega}{\sinh^2\gamma(\omega)}\mathrm{e}^{-n\gamma(\omega)}=M_1^2(0^+)+\Phi(n)\,, \tag{B30}$$

in $\Phi(n)$ is given by (50); in (B30), $M_1(0^+)$ is given as (B3).

As can be seen $M_1^2(0^+)=\lim_{N\to\infty}A_P=\lim_{L\to\infty}\lim_{N\to\infty}A_P$ for the case of $T$ far away from $T_c$ by comparing (B3) with (58) and (59), and, further, comparing (B29) and (B30) with (49) and (67), respectively, as can be seen that $\lim_{L\to\infty}\lim_{N\to\infty}\left\langle\sigma_{1,1}\sigma_{1+P,1}\right\rangle$ given as (49) and (67) are in accord with $S_{1,1}(P,0)$ given as (B29) and (B30) for $P=1,2,3,\cdots$ be small positive integer such that $\lim_{L\to\infty}(L-Q_1)=\infty$, respectively.

**B.3 The case of $n\left|1-\frac{T}{T_c}\right|>>1$**

In the case of $n\left|1-\frac{T}{T_c}\right|>>1$, the asymptotic expansion of $S_{1,1}(n,0)$ given as (B4) and (B5) becomes (VII.5.29) and (VII.6.11) in Ref.[1], respectively, i.e.,

$$S_{1,1}(n,0)\sim\frac{1}{2\sqrt{2\pi}}\frac{1-z_2^2}{z_1z_2^2}\frac{\sqrt{x_3+1}}{\sqrt{x_1+1}\sqrt{x_2+1}}\frac{1}{\alpha_2^{n+1}n^{3/2}}\left(1+O\left(\frac{1}{n}\right)\right)\qquad\text{for } T>T_c\,, \tag{B31}$$

$$S_{1,1}(n,0)\sim M_1^2\left(0^+\right)+\frac{1}{2\sqrt{2\pi}}\frac{1-z_2^2}{z_1z_2^2}\frac{\sqrt{|x_3+1|}}{\sqrt{x_1+1}\sqrt{x_2+1}}\frac{\alpha_2^{n-1}}{n^{3/2}}\left(1+O\left(\frac{1}{n}\right)\right)\qquad\text{for } T<T_c\,, \tag{B32}$$

in (B31) and (B32),

$$x_1=\frac{1+\alpha_1/\alpha_2}{1-\alpha_1/\alpha_2}\,,\qquad x_2=\frac{1+\alpha_1\alpha_2}{1-\alpha_1\alpha_2}\,,\qquad x_3=\frac{\alpha_2^2+1}{\alpha_2^2-1}\,. \tag{B33}$$

In terms of (B2), (B9) and (B33), (B31) and (B32) can be written in the form

$$S_{1,1}(n,0)\sim\frac{1}{\sqrt{\pi}}\left(\frac{\sinh 2K'}{\tanh K}\right)^{\frac{3}{2}}\frac{\mathrm{e}^{-2K'}\cosh K}{\sqrt{\sinh 2(K'-K)}}\frac{1}{\mathrm{e}^{2n(K'-K)}n^{3/2}}\left(1+O\left(\frac{1}{n}\right)\right)\qquad \text{for } T>T_{\mathrm{c}}\,, \tag{B34}$$

$$S_{1,1}(n,0)\sim M_1^2\left(0^+\right)+\frac{1}{\sqrt{\pi}}\left(\frac{\sinh 2K'}{\tanh K}\right)^{\frac{3}{2}}\frac{\mathrm{e}^{-2K'}\cosh K}{\sqrt{\sinh 2(K-K')}}\frac{1}{\mathrm{e}^{2n(K-K')}n^{3/2}}\left(1+O\left(\frac{1}{n}\right)\right)\quad \text{for } T<T_{\mathrm{c}}\,. \tag{B35}$$

In the case of $n=\left[\frac{L}{2}\right]\pm Q_2$ and $Q_2=0,1,2,3,\cdots$ be zero or small positive integer such that $\lim_{L\to\infty}\left(\left[\frac{L}{2}\right]\pm Q_2\right)=\infty$, as $L\to\infty$, $n\to\infty$, (B34) and (B35) become

$$\lim_{L\to\infty}S_{1,1}\left(\left[\frac{L}{2}\right]\pm Q_2,0\right)\sim\begin{cases}0\,, & \text{for } T>T_{\mathrm{c}}\,;\\ M_1^2\left(0^+\right), & \text{for } T<T_{\mathrm{c}}\,.\end{cases} \tag{B36}$$

As can be seen that, for $P=\left[\frac{L}{2}\right]\pm Q_2$, $\lim_{L\to\infty}\lim_{N\to\infty}\left\langle\sigma_{1,1}\sigma_{[L/2]\pm Q_2+1,1}\right\rangle$ given as (49) and (67) are in accord with $S_{1,1}\left(\left[\frac{L}{2}\right]\pm Q_2,0\right)$ given as (B36).

## References

[1] B. M. McCoy, T. T. Wu. *The Two-dimensional Ising Model.* Harvard University Press, 1973.

[2] B. M. McCoy. *Advanced Statistical Mechanics.* Oxford University Press, 2010.

[3] B. Kaufman. Crystal Statistics. II. Partition Function Evaluated by Spinor Analysis. *Phys. Rev.*, 1949, **76**(8), 1232-1243.

[4] Mei, T. An exact closed formula of a spin-spin correlation function of the two-dimensional Ising model with finite size. *Inter. J. Theo. Phys*. 2015, **54**, 3462-3489.

[5] Tao Mei. Exact expressions of spin-spin correlation functions of the two-dimensional rectangular Ising model on a finite lattice. *Entropy*, 2018, **20** (4), 277(1-12).

[6] V. V. Mangazeev, A. J. Guttmann. Form factor expansions in the 2D Ising model and Painlevé VI. Nuclear Physics B, **838**(3): 391-412 (2010).

[7] M. A. Rajabpour. Finite size corrections to scaling of the formation probabilities and the Casimir effect in the conformal field theories. Journal of Statistical Mechanics: Theory and Experiment, (2016) 123101.

[8] E. Granet, M. Fagotti, F. H. L. Essler. Finite temperature and quench dynamics in the Transverse Field Ising Model from form factor expansions. SciPost Phys. 9, 033 (2020).

[9] P. Di Francesco, P. Mathieu, David. Sénéchal. *Conformal Field Theory*. New York: Springer, 1997.